\documentclass[11pt,a4paper]{article}
\usepackage{jcappub}

\usepackage{graphicx}
\usepackage{amsmath}
\usepackage{epsfig,multicol,bbm}

\newcommand{\be}{\begin{equation}}
\newcommand{\ee}{\end{equation}}
\newcommand{\bea}{\begin{eqnarray}}
\newcommand{\eea}{\end{eqnarray}}

\newcommand{\mc}{\mathcal}

\newcommand{\beqa}{\begin{eqnarray}}
\newcommand{\eeqa}{\end{eqnarray}}

\newcommand{\vo}{\mathcal{V}}

\newcommand{\nn}{\nonumber}

\newcommand\fverb{\setbox\fverbbox=\hbox\bgroup\verb}
\newcommand\fverbdo{\egroup\medskip\noindent
			\fbox{\unhbox\fverbbox}\ }
\newcommand\fverbit{\egroup\item[\fbox{\unhbox\fverbbox}]}
\newbox\fverbbox

\author[a,b]{Michele Cicoli,}
\author[c]{Francisco G. Pedro,}
\author[d]{Gianmassimo Tasinato}

\affiliation[a]{Abdus Salam ICTP, Strada Costiera 11, Trieste 34014, Italy}
\affiliation[b]{INFN, Sezione di Trieste, Italy}
\emailAdd{mcicoli@ictp.it}
\affiliation[c]{Rudolf Peierls Centre for Theoretical Physics, University of Oxford,\\ 1 Keble Road, Oxford, OX1 3NP, UK}
\emailAdd{f.pedro1@physics.ox.ac.uk}
\affiliation[d]{Institute of Cosmology \& Gravitation, University of Portsmouth, \\
Dennis Sciama Building, Portsmouth, PO1 3FX, United Kingdom}
\emailAdd{gianmassimo.tasinato@port.ac.uk}

\title{Natural Quintessence in String Theory}

\abstract{We introduce a natural model of quintessence in string theory
where the light rolling scalar is radiatively stable
and couples to Standard Model matter with weaker-than-Planckian strength.
The model is embedded in an anisotropic type IIB compactification
with two exponentially large extra dimensions and TeV-scale gravity.
The bulk turns out to be nearly supersymmetric since the scale of the gravitino mass
is of the order of the observed value of the cosmological constant.
The quintessence field is a modulus
parameterising the size of an internal four-cycle which naturally develops a potential of the order (gravitino mass)$^4$,
leading to a small dark energy scale without tunings.
The mass of the quintessence field is also radiatively stable since it is protected by supersymmetry in the bulk.
Moreover, this light scalar couples to ordinary matter via its mixing with the volume mode.
Due to the fact that the quintessence field is a flat direction at leading order, this mixing
is very small, resulting in a suppressed coupling to Standard Model particles
which avoids stringent fifth-force constraints.
On the other hand, if dark matter is realised in terms of Kaluza-Klein states,
unsuppressed couplings between dark energy and dark matter can emerge, leading to a scenario of coupled quintessence within string theory.
We study the dynamics of quintessence in our set-up, showing that its main features make it compatible with observations.}

\begin{document}
\maketitle

\section{Introduction}

Present cosmological observations indicate that our Universe is currently accelerating.
This accelerated expansion is driven by a mysterious form of energy density with negative pressure, dubbed `dark energy',
that accounts for about 70$\%$ of the total energy density of the Universe.
The major part of the remaining 30$\%$ consists of cold non-baryonic dark matter, whose microscopic origin is not known,
although it is likely to be associated with new particles as supersymmetric partners, axions, string moduli or hidden sector degrees of freedom.
Finally, ordinary baryonic matter accounts for only a few$\%$ of the total energy density.

The microscopic origin of dark energy is even less clear than that of dark matter. Many proposals have been pushed
forward for explaining present day acceleration (see \cite{QuintReview} for a comprehensive review).
We will focus here on quintessence models, in which the dynamics of scalar fields drives the present cosmological acceleration,
as opposed to the simplest scenario in which the cosmological
energy density is dominated today by a pure cosmological constant.

Both of these proposals, a rolling quintessence field and a non-zero cosmological constant,
have their own virtues and shortcomings:
\begin{itemize}
\item \emph{Cosmological Constant Problem}: If dark energy is given by a non-zero cosmological constant,
in order to match the current cosmological observations,
its value needs to be of the order $\langle V \rangle = \Lambda^4 \sim \text{meV}^4 \sim 10^{-120}\,M_P^4$.
Whereas, if dark energy is provided by a quintessence model, in which
a late-time acceleration is characterised by a scalar $\phi$ with kinetic energy smaller than its potential energy,
the corresponding scalar mass has to be of the order
$m \sim 10^{-33}\,$eV. The two values of $\Lambda$ and $m$ are related to each other since:
\be
m^2\sim V_{\phi\phi} \sim \frac{V}{\phi^2} \lesssim \frac{V}{M_P^2} \sim 10^{-120} M_P^2 \sim \left( 10^{-33}\,\text{eV} \right)^2\,.
\ee
The condition $\phi \gtrsim M_P$ follows from the Friedmann equation $3 M_P^2 H^2 = (\dot{\phi}^2/2 + V ) \simeq V$
and the equation of motion for $\phi$, $0 = \ddot{\phi} + 3 H \dot{\phi} + V_\phi \simeq 3 H \dot{\phi} + V_\phi$.
In fact, for $\dot{\phi}^2\lesssim V$, we have:
\be
\phi \sim \frac{V}{V_\phi} \sim \frac{V}{H \dot{\phi}} \gtrsim M_P\,.
\ee
The required smallness of these quantities ($\Lambda$ or $m$, depending on the scenario under consideration)
represents one of the biggest mysteries of modern physics since a na\"ive estimate of
the vacuum energy would yield a result many orders of magnitude larger than the observed one:
$\langle V \rangle \sim M_P^4$ for non-supersymmetric theories and
$\langle V \rangle \sim \text{TeV}^4 \sim 10^{-60}\,M_P^4$ for theories with TeV-scale supersymmetry.
In the quintessence case, one can assume that the vacuum energy is exactly zero (as forced by symmetry arguments,
for example),  but one should still explain how to obtain such a small mass for the rolling scalar.
Indeed, scalar masses are notoriously hard to keep from getting large contributions when integrating out
ultra-violet physics running in the loops. This is the issue of \emph{technical naturalness},  which
plays a crucial r\^ole  in the determination of the Higgs mass and the solution of the famous $\eta$-problem
in slow-roll inflation. However, here the problem is more severe than in inflation
since one has to integrate out all the modes down to the tiny cosmological constant scale $\Lambda \sim \mc{O}(\text{meV})$,
instead of  the inflationary scale $M_{\rm inf} \sim \mc{O}(M_{\rm GUT})$.

\item \emph{Coincidence Problem}: Observations indicate that the energy densities of
dark matter and dark energy are of the same order of magnitude at present cosmological epochs.
The reason why different forms of matter have similar properties today
is a puzzle which goes under the name of `coincidence problem'.
Dark energy models based on a non-zero, pure cosmological constant have little to say about it, but quintessence models
can in principle provide answers to this problem. The idea is to consider `tracking' solutions,
in which the non-trivial dynamics of the scalar field leads
its energy density to mimic radiation during the radiation-dominated era.
The transition between radiation to matter domination, then,
triggers quintessence towards assuming characteristics
that match the observed behavior of dark energy.

\item \emph{Fifth-force constraints}:
Despite providing a potential understanding of the present ratios of energy densities via tracking solutions,
quintessence models are generically plagued by the problem of long-range unobserved fifth-forces
which would be mediated by such a light scalar field \cite{Kolda:1998wq}.
The present mass bounds for scalars with Planck strength couplings to matter are $m\gtrsim \mc{O}(\text{meV})$ \cite{Adelberger:2003zx}.
Furthermore, a rolling scalar would lead to a time variation of the constants of Nature, while
these do not appear to drastically change between Big Bang Nucleosynthesis (BBN) and the present epoch.
The standard solution to this problem involves pseudo Goldstone bosons with shift symmetries \cite{Carroll:1998zi}
since these pseudoscalars give rise to spin-dependent couplings which cannot be probed in
fifth-force experiments that are insensitive to the polarization.
Moreover, the shift symmetry forbids higher dimensional operators, rendering the quintessence field
stable against radiative corrections from UV physics.
On the other hand, it is much harder to satisfy the current constraints in models where the quintessence field is a scalar.
\end{itemize}

Most of the attempts to realise quintessence within string theory rely on
axions and shift symmetries \cite{Choi:1999xn,Kaloper:2008qs,Panda:2010uq,Gupta:2011yj}.
In this paper we shall, however, take a different approach, and build
the first string model where the quintessence field is a scalar parameterising
the size of an internal four-cycle. We shall focus on LARGE Volume Scenarios (LVS)
which emerge naturally in type IIB flux compactifications.
The set-up of this model represents an interesting variation of the one derived in \cite{Cicoli:2011ct}
for inflationary purposes. In fact, as in \cite{Cicoli:2011ct}, we focus on the recent string embedding \cite{ADDstrings}
of the Supersymmetric Large Extra Dimension (SLED) proposal \cite{SLED}.
The field which was playing the r\^ole of the inflaton in \cite{Cicoli:2011ct} will now behaves as
quintessence.

In our new model, quintessence is driven by a closed string modulus
with a naturally small mass $m$ protected against loop corrections.
The mechanism for obtaining such a light scalar whose mass is radiative stable is similar
to the one of \cite{Albrecht:2001xt} and we shall outline it in section \ref{NatQuint}.
Contrary to the model of \cite{Albrecht:2001xt} where the quintessence field was the overall volume mode,
in our case quintessence is driven by a fibre modulus that at leading order
neither develops a potential nor couples to Standard Model (SM) matter.
The potential for the quintessence field arises only at subleading order
by means of non-perturbative and string loop effects, that also induce suppressed,
weaker-than-Planckian couplings with ordinary matter. Thus, our model is able to avoid
fifth-force constraints.
An intuitive explanation for this non-standard behaviour for the quintessence scalar will be given in section \ref{NatQuint},
while in sections \ref{QuintPot} and \ref{QuintCoupl} we will develop our arguments at a more technical level.

The dynamics of the quintessence-dark matter system will then be analysed in section \ref{Dynamics} starting
from the standard decoupled case and then moving to the
more complicated coupled case where we shall also explain
how our stringy set-up can allow for unsuppressed couplings of the quintessence field
with dark matter degrees of freedom. Hence we will suggest a realisation of a coupled quintessence-dark matter model
inspired by string theory.
Finally we will conclude in section \ref{Conclusions}.

\section{Natural quintessence from anisotropic compactifications}
\label{NatQuint}

In this section, we shall outline in some detail the ideas behind the construction of our
quintessence model, before developing the technical elements of our set-up in the next sections.

\subsection{Naturally light quintessence field}
\label{LightQuint}

In our approach, a naturally small mass for the quintessence field
is generated by means of a string realisation
of the so-called Supersymmetric Large Extra Dimension (SLED)
scenario (see \cite{Burgess:2004ib} for a review). In its original form,
this is a brane-world model with two large extra dimensions, in which supersymmetry is preserved
in the bulk, whereas it is only non-linearly realised on a Standard Model codimension-two brane.
Its main attractive feature is the possibility to tie the discussion  of the hierarchy problem
to that of the cosmological constant problem.
More precisely, the hierarchy problem is addressed  by the presence of large extra dimensions which lower
the fundamental gravity scale down to the TeV-scale; at the same time,
the observed value of the cosmological constant is associated with the low supersymmetry breaking scale in the bulk.
These two arguments are tied up since simple considerations based on
dimensional reduction yield the following relation among the fundamental $(4+d)$-dimensional
gravity scale $M_*$ and the gravitino mass $m_{3/2}$:
\be
m_{3/2} \sim \left(\frac{M_*}{M_P}\right)^2 M_P\,.
\label{m32}
\ee
Setting $M_* \sim \mc{O}(\text{TeV}) \sim 10^{-15} M_P$, the gravitino mass turns out to be very small, of the order
of the dark energy scale
$m_{3/2}\sim\Lambda^{1/4} \sim \mc{O}(\text{meV}) \sim 10^{-30} M_P$. This implies that the bulk is nearly supersymmetric, given that
supersymmetry is broken only at the dark energy scale. Quantum corrections
to the vacuum energy from loops of bulk fields scale as:
\be
V_{\rm loop} \sim \text{Str}(M^2)\, \Lambda_{\rm UV}^2 \sim m_{3/2}^2\, M_{\rm KK}^2\,,
\label{CWV}
\ee
where $\text{Str}(M^2) \sim m_{3/2}^2$ and the cut-off is taken to be the scale at which a pure  4D description
ceases to be valid: $\Lambda_{\rm UV}\sim M_{\rm KK}$. Given that the Kaluza-Klein scale is given by:
\be
M_{\rm KK} \sim \left(\frac{m_{3/2}}{M_P}\right)^{1/2+1/d}\,M_P\,,
\ee
this is of the same order of $m_{3/2}$ for $d=2$. This is also the case that for
TeV-scale fundamental gravity leads to micron-sized extra dimensions
which are at the edge of detectability in experiments that look for deviations from Newton's law.
Hence in the case of two large extra dimensions, the loop generated potential (\ref{CWV})
gives the correct observed order of magnitude of the cosmological constant since it scales as
$V_{\rm loop} \sim m_{3/2}^4 \sim 10^{-120} M_P$. Notice that for energies above the cut-off $M_{\rm KK}$,
the theory is effectively 6D and supersymmetric, and so the contributions to the vacuum energy
from loops of bosons and fermions cancel among each other.

In order to have a viable and complete solution of the cosmological constant problem, one has to make sure that
(\ref{CWV}) is the leading order contribution to the vacuum energy. In general, one would expect
larger contributions coming directly from the tension of the Standard Model brane,
of the order $T \sim \mc{O}(\text{TeV})^4$, and from the tree-level part of the
bulk potential corresponding to the curvature of the extra dimensions induced by the presence of the TeV brane.
This is a delicate issue (see e.g. \cite{Nilles:2003km}): recent progresses seem to suggest
that consistent scenarios can be built along these lines by means of suitable couplings between bulk fields
and the SM brane \cite{Burgess:2011mt}. Since the rest of our discussion does not rely specifically on these
details, we will not develop this issue any further in this paper.

\smallskip

The SLED framework opens the possibility to obtain a natural model of quintessence.
In fact, in 6D SLED models, the radion mode $r$ develops a potential via loops of bulk fields (Casimir energy)
which scale as (\ref{CWV}), and so give it a mass of the order \cite{Albrecht:2001xt}:
\be
m^2_r \sim \frac{V_{\rm loop}}{M_P^2} \sim \frac{m_{3/2}^4}{M_P^2} \sim \left(10^{-33}\,\text{eV}\right)^2\,.
\label{mr2}
\ee
This is the perfect scale for quintessence, since it is the typical mass scale expected for a quintessential
scalar; for the arguments we outlined above,
this small mass is \emph{technically natural} since it is radiatively stable thanks to the unbroken supersymmetry in the bulk
(see \cite{Albrecht:2001xt} for more details).

\smallskip

In the quintessence model based on the 6D SLED
scenario, the radion mode, being a scalar instead of a pseudo-scalar
with Planck strength coupling to ordinary matter, would mediate long range fifth-forces.
Moreover its cosmological evolution would lead to a time dependence of Newton's constant.
The authors of \cite{Albrecht:2001xt} propose  to cure this problem exploiting the fact that
the couplings of the radion to matter are field dependent,  approaching  small values for appropriate
cosmological evolutions of the radion field. The strongest constraint comes from the requirement
that the Newton's constant does not change that much from BBN till today.

We will now argue that a natural, automatic
solution to the fifth-force problem emerges when embedding this quintessence scenario in string theory.
A recent paper \cite{ADDstrings} discusses how to embed these 6D SLED scenarios within type IIB flux compactifications.
The simplest Calabi-Yau three-fold that allows such a derivation has a fibred structure with a
4D K3 or $T^4$ fibre over a 2D $\mathbb{P}^1$ base. This structure allows to take a very anisotropic limit where
the base is exponentially larger than fibre, resulting in an intermediate 6D effective field theory when
compactifying 10D type IIB string theory down to 4D. The authors of \cite{ADDstrings} show how to stabilise
the geometric moduli in such a way to obtain the above mentioned anisotropic limit within the
framework of the LARGE Volume Scenario (LVS) \cite{LVS}. In this framework, exponentially large extra dimensions
emerge from a moduli stabilisation mechanism that exploits a combination of non-perturbative and $\alpha'$ effects.
This set-up provides the perfect arena for building a SLED-motivated quintessential scenario in IIB string theory:
in the next subsection we will outline why and how it automatically solves the problem of fifth-forces.

\subsection{Naturally decoupled quintessence field}
\label{subscoupl}

The explicit string embedding reveals that the 6D effective field theory is  far more richer than the one
considered in \cite{Albrecht:2001xt}, due to the presence of additional closed string moduli on top of the radion (that in the IIB embedding
corresponds to the volume mode).
Moreover, besides the equivalent of the loop corrections (\ref{CWV}), both higher-derivative contributions
($\alpha'$ corrections) and non-perturbative effects play a crucial r\^ole in determining the bulk action.
These additional features of the 6D effective field theory,
obtained as the low-energy limit of an anisotropic fibred compactification,
might at first glance seem a source of new complications
for getting a viable quintessence model.
However, a deeper analysis reveals that the string embedding actually  provides an elegant
quintessence model,  with a natural solution to the fifth-force problem.

The minimal scenario we consider in this work involves three K\"ahler moduli, which parameterise the size in string units
of internal four-cycles, and a fibered Calabi-Yau volume of the form \cite{FibredCY}:
\be
\vo = t_b\,\tau_f-\tau_s^{3/2}\,.
\label{CYvolume}
\ee
The fields defining the volume have the following r\^oles and
geometrical interpretation:
\begin{itemize}
\item $\tau_s$ is a del Pezzo divisor supporting non-perturbative effects of the form $W_{\rm np} = A\,e^{-a T_s}$,
where $\tau_s = {\rm Re}(T_s)$ and $A$ and $a$ are $\tau_s$-independent constants. These effects fix $\tau_s$ at `small' values
but still in a geometric regime above the string scale: $\langle\tau_s\rangle \sim \mc{O}(10)$.

\item The overall volume mode $\vo \simeq t_b  \tau_f$ is a modulus
fixed at exponentially large values, $\langle\vo\rangle\sim e^{1/g_s} \gg 1$ for weak string coupling $g_s \lesssim 0.1$,
by the interplay of non-perturbative effects and $\alpha'$ corrections to the K\"ahler potential \cite{LVS}.
The size of the 2D $\mathbb{P}^1$
base is given by $t_b$ whereas the volume of the 4D K3 or $T^4$ fibre is given by $\tau_f$.
The value of $\vo$ plays a crucial r\^ole in determining the scales and coupling strengths of our model, explaining in a natural
way why the size of dark energy is so small.

\item The fibre modulus $\tau_f$ plays the r\^ole of the quintessence field.
It develops a potential only a subleading order due to poly-instanton corrections to
the superpotential generated by an E3-instanton wrapping this non-rigid divisor: $W_{\rm np} = A\,e^{-a\left(T_s + C e^{-2\pi T_f}\right)}$,
where again $\tau_f = {\rm Re}(T_f)$ and $C$ is a $\tau_f$-independent constant.
These tiny effects fix the fibre modulus `small',
i.e. at the same order of magnitude of $\tau_s$: $\langle\tau_f\rangle \sim \langle\tau_s\rangle$. This gives rise to a very anisotropic
compactification with $t_b$ exponentially larger than $\sqrt{\tau_f}$.
\end{itemize}
None of these three four-cycles can support the visible sector where supersymmetry is non-linearly 
realised via a proper configuration of branes wrapped around internal divisors.\footnote{We need to focus on a 
non-linear realisation of supersymmetry where the low-energy effective field theory contains just the Standard Model without any 
superpartner since models with TeV-scale strings predict $m_{3/2}\sim$ meV, and so gravity mediation would give rise to soft terms 
of this unacceptably small order of magnitude.}
In fact, the Standard Model cannot be localised neither on $\tau_s$ nor on $\tau_f$ due to the tension between chirality
and the generation of non-perturbative effects \cite{Blumenhagen:2007sm}. Moreover, the four-cycle containing the base, $\tau_b \simeq t_b \sqrt{\tau_f}$,
would give rise to an exponentially small gauge coupling $g^{-2}\simeq \tau_b \gg 1$. Thus we need to add a fourth
K\"ahler modulus for the Standard Model, $\tau_{\rm SM}$. This divisor has to be rigid in order to avoid undesired matter
in the adjoint representation and has to be fixed at small values in order to get a correct value of the gauge coupling.
This can be done by exploiting D-terms and string loop corrections to the K\"ahler potential \cite{Cicoli:2011qg}.
Therefore $\tau_{\rm SM}$ has the same features as $\tau_s$ apart from the fact that one cycle is fixed perturbatively whereas the other
non-perturbatively. Hence, without loss of generality, we shall just restrict ourselves to the simplest case with three moduli
having in mind that the modulus $\tau_s$ can represent a generic rigid and `small' divisor
which can be either $\tau_{\rm SM}$ or the one supporting the non-perturbative effects.

The structure of the scalar potential for the bulk fields (trading $t_b$ for $\vo$) is the following:
\be
V = V_{\rm np}\left( \vo, \tau_s \right) +  V_{\alpha'}\left( \vo, \tau_s \right)
+  V_{\rm poly}\left( \vo, \tau_s, \tau_f \right) +  V_{\rm loop}\left( \vo, \tau_s, \tau_f \right)\,,
\label{Vtot}
\ee
where the leading order effects are given by the non-perturbative and $\alpha'$ potentials, $V_{\rm np}$ and $V_{\alpha'}$,
which depend only on $\vo$ and $\tau_s$ and scale as:
\be
V_{\rm lead} = V_{\rm np}\left( \vo,\tau_s \right) +  V_{\alpha'}\left( \vo,\tau_s \right)
\sim \mc{O}\left(m_{3/2}^3 M_P\right) \,.
\label{Vlead}
\ee
The previous potential $V_{\rm lead}$ determines the properties of the volume and the modulus $\tau_s$ at leading
order in an inverse volume expansion (since the volume is exponentially large, its inverse represents an ideal expansion
parameter). This implies that the basic features of SM physics, like gauge and Yukawa couplings,
are determined at leading order by the VEV of $\tau_s$.
Instead, in our string construction,
the fibre modulus $\tau_f$ develops a potential only at subleading order due to poly-instanton effects and
loops of closed strings propagating in the bulk. The subleading potential  scales as (\ref{CWV}):\footnote{We do not include the contribution
coming from loops of open string fields since by construction no D7-brane stack wraps either the fibre or the base divisor.
Hence open strings are localised far away in the Calabi-Yau, and so do not develop any potential for $\tau_1$.
If that were the case, the internal manifold would have a isotropic shape since $\tau_1$ would also be fixed at exponentially large values:
$\langle\tau_1\rangle \simeq g_s^{4/3}\,\langle\vo\rangle^{2/3}\gg 1$.}
\be
V_{\rm sub} = V_{\rm poly}\left( \vo, \tau_s, \tau_f  \right) +  V_{\rm loop}\left(  \vo, \tau_s, \tau_f  \right)\sim \mc{O}\left(m_{3/2}^4 \right)\,.
\label{Vsub}
\ee
If the tension of the Standard Model brane is compensated by the brane back-reaction,
as in the 6D SLED scenario, the leading order contribution to the vacuum energy is given by (\ref{Vlead})
which does not reproduce the correct value of the cosmological constant for $m_{3/2}\sim \mc{O}(\text{meV})$
obtained from (\ref{m32}) setting the string scale $M_s = M_*$ of the order the TeV scale.
However the potential (\ref{Vlead}) gives rise to a SUSY breaking AdS vacuum, which still needs to be uplifted to a dS solution.
This can be successfully accomplished by means of one of the many mechanisms known in the literature
(tension of anti branes at the tip of warped throats  \cite{Kachru:2003aw}, D-terms from
magnetised branes  \cite{bkq} or dilaton-dependent non-perturbative effects \cite{Cicoli:2012fh}).
The leading order contribution to the vacuum energy would then be given by the subleading potential (\ref{Vsub}) which
now naturally gives rise to the observed size of dark energy,
since it scales as the bulk loop potential of 6D SLED models (see (\ref{CWV})).

A key difference  between our scenarios and SLED constructions
is the fact that the volume mode can no longer be used as the quintessence field, since it develops a potential at
order $m_{3/2}^3 M_P$, and so the corresponding mode is too heavy to drive quintessence:
\be
m_{\vo}^2 \sim \frac{V_{\rm lead}}{M_P^2} \sim m_{3/2}^2\left(\frac{m_{3/2}}{M_P}\right) \sim \left(10^{-18} \,\text{eV}\right)^2\,.
\label{volmass}
\ee
On the other hand, the fibre modulus $\tau_f$ does {\it not} appear in the expression (\ref{Vlead}) for $V_{\rm lead}$,
and so its mass has the perfect quintessential value which can be estimated from the subleading potential (\ref{Vsub}):
\be
m_{\tau_f}^2 \sim \frac{V_{\rm sub}}{M_P^2} \sim m_{3/2}^2\left(\frac{m_{3/2}}{M_P}\right)^2 \sim \left(10^{-33} \,\text{eV}\right)^2\,.
\label{taufmass}
\ee
This expression is similar to (\ref{mr2}) which gives the mass of the radion mode of vanilla 6D SLED models.
Consequently the natural explanation of why the quintessence field is light and radiatively stable is, in practice, the same in
both scenarios.
However, the radion mode in the SLED proposal couples to ordinary matter with Planck strength, $1/M_P$. 
Instead, as discussed also in \cite{ADDstrings},
the fibre modulus has a weaker-than-Planckian coupling of the order $\epsilon/M_P$ 
where $\epsilon \sim m_{3/2}/M_P \ll 1$.
For this reason, this set-up is ideal for building a quintessence model since the rolling field
can safely evade all the current bounds from fifth-force experiments.

However, given that $\vo$ couples to ordinary matter as $1/M_P$, 
these bounds would not be evaded by the volume mode if it had a mass below the meV scale 
as estimated in (\ref{volmass}).
Nonetheless, supersymmetry is only non-linearly realised on the SM brane, 
and so the moduli masses (\ref{volmass}) and (\ref{taufmass}) might get 
large radiative corrections from loops of open strings.\footnote{These loops of brane fields contribute to the brane tension 
which has to be cancelled by the brane back-reaction. However this cancellation concerns just the vacuum energy, 
i.e. the value of the potential at the minimum.} As estimated in \cite{ADDstrings},
this is indeed the case for the volume mode whose mass receives a correction of the order:
\be
\delta m_{\vo} \simeq  \frac{M_s^2}{M_P} \simeq m_{3/2} \sim \mc{O}\left(\text{meV}\right)\,,
\ee
which renders this modulus still compatible with fifth-force bounds. 
We stress that the mass of the quintessence field does not get lifted by loops of brane fields 
due to the weakness of its coupling. In fact, the correction estimated in \cite{ADDstrings} scales as:
\be
\delta m_{\tau_f} \simeq  \epsilon\,\frac{M_s^2}{M_P} \simeq \epsilon \,m_{3/2} \sim \mc{O}\left(10^{-33} \,\text{eV}\right)\,,
\ee
and so does not lift the mass (\ref{taufmass}).

The intuitive reason of the weakness of the fibre modulus coupling to ordinary matter is the fact that the leading order potential $V_{\rm lead}$
does not depend on $\tau_f$ (see (\ref{Vlead})), and so there is no leading order mixing between $\tau_f$ and the volume mode $\vo$.
The mixing between $\tau_f$ and $\vo$ arises only at subleading order at the level of $V_{\rm sub}$ (see (\ref{Vsub})).
Therefore the order of magnitude of this small mixing, $\epsilon \ll 1$, is weighted by the ratio between $V_{\rm sub}$ and $V_{\rm lead}$:
\be
\epsilon \sim \frac{V_{\rm sub}}{V_{\rm lead}}\sim \mc{O}\left(\frac{m_{3/2}}{M_P}\right)\,.
\ee
Given that the volume mode couples to ordinary matter as $1/M_P$ and
the coupling of the fibre modulus to SM particles is induced by the mixing between $\tau_f$ and $\vo$, the coupling
of the fibre modulus has to be suppressed with respect to the one of the volume mode by the small mixing parameter $\epsilon$, explaining
the weaker-than-Planckian nature of this interaction.
In other words, the weakness of this coupling is reflecting the fact that $\tau_f$ is a flat direction at leading order
and does not couple with ordinary matter at that level.
On the other hand, the quintessence field has the opportunity to couple with Planckian strength to massive degrees of freedom
(as stable 6D Kaluza-Klein states) that can provide a model of dark matter.
Consequently, our set-up can lead to a stringy realisation of an interacting dark energy-dark matter system.

We shall make this analysis more precise in section \ref{QuintCoupl} where we will compute the moduli couplings to Standard Model particles.
These are extracted from the moduli dependence of the kinetic and mass terms of gauge bosons and fermions, using the
canonical normalisation carried out in appendix \ref{AppendixA}.

\smallskip

After this outline of the properties of our quintessence scenario, we pass to develop the technical details.

\section{Explicit form of the quintessence potential}
\label{QuintPot}

In this section, we start developing our set-up by discussing the features of the potential that drives quintessence.
As explained in section \ref{NatQuint},
we work within the framework of the LARGE Volume Scenario of type IIB string theory \cite{LVS}
where the volume is fixed at exponentially large values by combining $\alpha'$ corrections to the K\"ahler potential $K$
and non-perturbative contributions to the superpotential $W$.

As explained in subsection \ref{subscoupl}, we consider compactifications where the volume of the Calabi-Yau
manifold has a fibered structure and is given by (\ref{CYvolume}).
With this choice of internal manifold, the standard LVS framework stabilises only two of the three K\"ahler moduli.
The remaining flat direction, which can be identified with the fibre modulus $\tau_f$,
is lifted either by string loop corrections to $K$ \cite{StringLoops}
or by poly-instanton effects \cite{ADDstrings, Cicoli:2011ct}. Here we focus our attention on the latter possibility,
and investigate if the resulting potential for the fibre modulus $\tau_f$
allows us to provide a string theory realisation of quintessence.
The analysis closely follows that of our previous paper \cite{Cicoli:2011ct},
in which we studied a similar scalar potential in the context of an inflationary model. For this reason,
we will only sketch the derivation and the properties of the scalar potential here,
referring the reader to \cite{Cicoli:2011ct} for more details.

It was shown in \cite{Blumenhagen:2008ji} that there are instantons which do not give rise to a single contribution to the superpotential but, due to the presence of extra fermionic zero-modes, they correct the action of another instanton wrapping a different internal cycle. These poly-instantons generate corrections to the superpotential of the form $W_{\rm np}\sim e^{-\left(T_i+e^{-T_j}\right)} $. These double exponential terms were used in \cite{ADDstrings} to provide a string realisation of 6D SLED scenarios and, in \cite{Cicoli:2011ct}, to provide a controlled model of inflation in the K\"ahler moduli sector.

In what follows we assume the following brane set-up: the cycle $\tau_s$ is wrapped by a stack
of magnetised D7-branes where appropriate gauge fluxes give rise to two different gauge groups
that separately undergo gaugino condensation. This gives rise to a race-track superpotential. Due to the existence of a Euclidean D3-instanton wrapping the fibre $\tau_f$, this racetrack superpotential acquires further corrections. Following \cite{Blumenhagen:2008ji}, the gauge kinetic functions of the two condensing gauge groups receive non-perturbative corrections, resulting in a poly-instanton corrected superpotential of the form:
\be
W=W_0+A \,e^{-a \left(T_s+C_1 e^{-2 \pi T_f}\right)} -B \,e^{-b \left(T_s+C_2 e^{-2\pi T_f}\right)}\,,
\label{eq:Wpoly}
\ee
Here, $A$ and $B$ are threshold effects, while $a=2 \pi/n_a$ and $b=2\pi/n_b$ with $n_{a, b}$ natural numbers.
$C_1$ and $C_2$ are constants associated with the effects of the D3-instanton wrapping $\tau_f$. 
On the other hand, the $\alpha'$ corrected K\"ahler potential reads: 
\be
K=K_0+K_{\alpha'}\simeq -2\ln\vo-\frac{\xi}{g_s^{3/2}\vo}\,,
\label{Kpot}
\ee
where $\xi$ is a positive topological quantity.

The scalar potential computed from eqs. (\ref{eq:Wpoly}) and (\ref{Kpot}) can be written as:
\be
V=V_{\mc{O}(\mc{V}^{-3})}\left(\tau_s,\vo\right)+V_{\mc{O}(\mc{V}^{-3-p})}\left(\tau_f,\tau_s,\vo\right)\,,
\ee
where the first term scales with the volume as ${\mc V}^{-3}$, while the other as ${\mc V}^{-3-p}$, with $p$ a positive,  order one parameter.
If $\vo$ is large, the second piece is  much suppressed with respect to the first one: the previous equation reproduces
the split between leading and subleading
potentials we discussed in section \ref{NatQuint} (see eq. (\ref{Vtot})).
The leading contribution to the scalar potential, $V_{\mc{O}(\mc{V}^{-3})}$,
is the standard LVS potential that stabilises the blow-up modulus $\tau_s$ and the volume mode $\vo$:
\be
\begin{split}
V_{\mc{O}(\mc{V}^{-3})}=&\frac{8 \sqrt{\tau_s}\left(A^2 a^2 e^{-2 a \tau_s}+B^2 b^2 e^{-2 b \tau_s} -2A B \, a b\,
 e^{- (a +b)\tau_s}\right)}{3 \mc{V}}\\
&+\frac{4 W_0 \tau_s \left(A a e^{- a \tau_s}-B b e^{- b \tau_s}\right)}{\mc{V}^2}+\frac{3 \xi W_0^2}{4 \,g_s^{3/2}\,\mc{V}^3}\,.
\end{split}
\label{eq:Vleading}
\ee
The subleading $V_{\mc{O}(\mc{V}^{-3-p})}$ term depends on all three directions of K\"ahler moduli space
and generates a minimum for the fibre modulus $\tau_f$:
\be
\begin{split}
V_{\mc{O}(\mc{V}^{-3-p})}=&-e^{-2 \pi \tau_f}\left\{ -\frac{16 C_1\sqrt{\tau_s}e^{-2 b\tau_s}}{3 \mc{V}}[Z^2 a-Z B (a-b)]\right.\\
&\left.+\frac{4 C_1 W_0 e^{-b\tau_s}}{\mc{V}^2}[2 \pi\,Z  \tau_f+Za\tau_s-B b(a-b)\tau_s]\right.\\
&\left.+n\left[\frac{-16 B b^2 \sqrt{\tau_s}e^{-2 b \tau_s}Z}{3\mc{V}}+\frac{4 W_0 B b e^{-b\tau_s}}{\mc{V}^2}(b \tau_s+2 \pi \tau_f)\right]\right\}\,,
\end{split}
\label{eq:Vsubleading0}
\ee
where $Z\equiv B b - A a e^{-(a-b) \tau_s}$ and $n\equiv C_2-C_1$.

Given that the mass spectrum in the K\"ahler moduli sector exhibits the following hierarchy
$m_s^2\sim \vo^{-2}\gg m_\vo^2 \sim \vo^{-3}\gg m_f^2 \sim \vo^{-3-p}$ with $p\sim\mc{O}(1)$ \cite{Cicoli:2011ct},
we set $\tau_s$ and $\vo$ to their minima determined (analytically) by minimising the potential (\ref{eq:Vleading}):
\be
\langle \tau_s \rangle^{3/2}=\frac{3\xi}{32\alpha\gamma f_1(1-f_1) g_s^{3/2}}\,,
\qquad \langle \vo \rangle= f_2 \,e^{b \langle\tau_s\rangle}\,.
\label{eq:LVSMmin}
\ee
Here we have defined:
\be
f_1\equiv \frac{\left(a-b\right) \tau_s B b+ Z (1-a\,\tau_s  )}{4 \left(a-b\right) \tau_s B b+Z(1-4 a\,\tau_s)}\,,
\qquad \text{and}\qquad f_2\equiv\frac{3 \alpha \gamma W_0 \sqrt{\langle\tau_s\rangle}}{ Z} f_1\,.
\ee
This allows us to reduce the analysis of quintessence to the study of a single field potential for $\tau_f$, the subleading
potential of eq. (\ref{eq:Vsubleading0}) where we substitute the VEVs of $\tau_s$ and $\vo$.
Following this procedure and performing the canonical normalisation of the fibre modulus (see appendix \ref{AppendixA}, in particular the discussion
around eq. (\ref{cant1})), the poly-instanton generated potential is written as:
\be
V_{\rm poly}= \frac{F_{\rm poly}}{\langle\vo\rangle^{3+p}} \left[1-\,e^{-c\left(e^{\frac{2}{\sqrt 3}
\hat \phi}-1\right)}\left(1+c\left(\,e^{\frac{2 \hat \phi}{\sqrt 3}}-1\right)\right)\right]\,,
\label{eq:VCanNorm}
\ee
where $\hat{\phi}$ is the field parameterising the shift of the canonically normalised fibre modulus from its minimum: $ \hat{\phi}\equiv \phi-\langle\phi\rangle=\sqrt{3}/2\ln (1+\hat{\tau}_f/\langle\tau_f\rangle)$. The constant $c$ is given by $c=2 \pi \langle \tau_f \rangle
= p b \langle \tau_s \rangle +1$.
$F_{\rm poly}$, on the other hand, is given in terms of the underlying parameters as:
\be
F_{\rm poly}= \left(\frac{3 \alpha W_0 \sqrt{\langle\tau_s\rangle}f_1}{Z} \right)^{p+1} 4 \,e^{-1} W_0 (C_1 Z- n B b).
\ee
Equation (\ref{eq:VCanNorm}) provides a first, extremely flat contribution to the quintessence potential we will be working  with.
Note that, as explained in section \ref{NatQuint}, we included in eq. (\ref{eq:VCanNorm}) an uplifting term,
$V_{\rm up}=F_{\rm poly}/\langle\vo\rangle^{3+p}$ so to set the minimum of the quintessence potential to zero.

\begin{figure}[h]
\begin{center}
\includegraphics[width=0.5\textwidth]{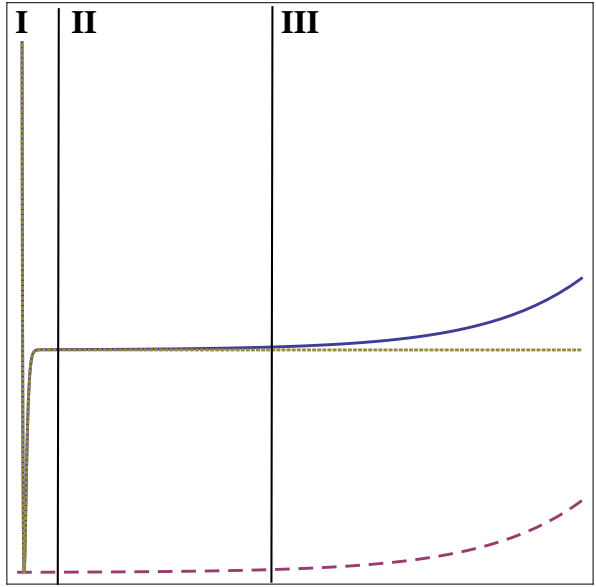}
\caption{Quintessence potential for the fibre modulus. The dotted line represents the poly-instanton generated potential, the dashed line the loop potential from UV physics and the full line the total potential for the canonically normalised fibre modulus.}
\label{fig:Potential}
\end{center}
\end{figure}

As pointed out in \cite{Cicoli:2011ct}, there will be further contributions to the fibre modulus potential,
in particular UV physics will add new $\phi$ dependent terms to the total potential.
These are loop contributions that we have to take into account, since they
can influence the dynamics of the quintessence field.
These contributions can be estimated via the Coleman-Weinberg potential and are
due to subleading $g_s$ corrections to the K\"ahler potential.
Supersymmetry reduces their size, and following the straightforward estimate performed in \cite{Cicoli:2011ct}, we obtain:
\be
V_{\rm loop} \simeq \frac{F_{\rm loop}}{\langle\vo\rangle^4}\,e^{\frac{2}{\sqrt 3}\hat \phi}\,,
\qquad\text{where }\qquad F_{\rm loop}= \frac{c}{2\pi}\left(g_s C_{\rm loop} W_0\right)^2\,.
\label{eq:VLoop}
\ee
The full potential for the fibre modulus that corresponds to our quintessence field is then given by:
\be
V_{\rm fibre}=V_{\rm poly}+V_{\rm loop},
\ee
which we plot in figure \ref{fig:Potential}. Given that the gravitino mass in terms of the overall volume is $m_{3/2}\sim M_P/\vo$, 
this potential for $p\sim\mc{O}(1)$ scales as $V_{\rm fibre}\sim \left(M_P/\vo\right)^4 \sim m_{3/2}^4$, reproducing the 
behaviour described in section \ref{NatQuint}.

For the study of the dynamics of the quintessence field it is convenient to divide the potential into three distinct regions:
\begin{enumerate}
\item The region I corresponds to the vicinity of the minimum where the potential is very steep;

\item The region II is the poly-instanton dominated part of the potential;

\item In region III the loop corrections are the dominant contribution to $V_{\rm fibre}$.
\end{enumerate}
Notice that, as outlined in section \ref{NatQuint}, we
have obtained an extremely flat quintessential potential (above all in region II) in which supersymmetry in the extra-dimensions allows
to tame the dangerous corrections that would spoil its flatness.
The aim of section \ref{Dynamics} of this paper is to investigate whether the dynamics of the fibre modulus rolling down regions II and III gives a good description of the observed late time acceleration of Universe.

Let us conclude this section determining the relevant scales in our set-up. As we explained,
observations constrain the dark-energy scale to be of the order $\rho_{DE} \sim 10^{-120} M_P^4$.
In order to use the fibre modulus as a dynamical model of dark-energy, this maps into the following condition on the amplitude
 of the potential:
\be\label{tihe}
\frac{F_{\rm poly}}{\langle\vo\rangle^{3+p}}\sim10^{-121}.
\ee
This scale can be achieved by considering compactifications with very large volumes,
as originally noted in \cite{ADDstrings}, in particular for $p\sim \mc{O}(1)$ this yields:
\be
\langle \vo\rangle\sim 10^{30}
\ee
in string units. We stress that,
 in our scenario, the VEV of the volume depends exponentially on the microscopic parameters of the model (see eq.
(\ref{eq:LVSMmin})), and such large values are easy to obtain. So, once the minimum of the
quintessence potential is set to zero, the tiny height of the potential, eq. (\ref{tihe}), is obtained straightforwardly!
Notice that values of the volume of the order $\vo\sim 10^{30}$ correspond to a string scale of the order the TeV-scale,
$M_s \sim M_P / \sqrt{\vo} \sim \mc{O}\left(\text{TeV}\right)$, and a gravitino mass of the order the cosmological constant scale,
$m_{3/2} \sim M_P / \vo \sim \mc{O}\left(\text{meV}\right)$, according to the discussion of section \ref{NatQuint}.
Such a low-value of the fundamental gravity scale poses a severe problem for obtaining the original period of inflation
since both the amount of density perturbations and the resulting reheating temperature would be too low.
For example, the inflationary model developed in \cite{Cicoli:2011ct} requires values of the volume of the order $\vo \sim 10^3$
corresponding to a GUT-scale $M_s$. The general idea that we have in mind to overcome this problem is to use the
volume mode as the inflaton following the work of \cite{Conlon:2008cj}. Thus the volume can take smaller values
of the order $\vo \sim 10^3$ in the early Universe,
and then roll to values of the order $\vo \sim 10^{30}$ after the end of inflation,
where the only field left over to roll down its minimum is the fibre modulus $\tau_f$ which drives quintessence.

A very large volume, apart from dynamically describing the smallness
of the observed size of dark energy, has also the virtue to suppress fifth-force couplings between the quintessence
field and SM matter, and to enlarge the very flat region II in the potential $V_{\rm fibre}$.
In the next sections, we will discuss these features in detail.

\section{Quintessence coupling to matter}
\label{QuintCoupl}

The first step in the computation of the quintessence couplings to ordinary matter is the moduli canonical normalisation,
which we perform in detail in appendix \ref{AppendixA}. The expression of the original moduli $\tau_f$, $\vo$, and $\tau_s$
in terms of the corresponding canonically normalised fields $\phi$, $\chi_{\vo}$ and $\chi_s$ can be obtained explicitly
at leading order in an inverse volume expansion. Considering all the moduli in the vicinity of their minimum,
the canonical normalisation of the Standard Model modulus represented by $\tau_s$, takes the form (see eq. (\ref{expt3})):
\be
\frac{\delta \tau_s}{\langle\tau_s\rangle}
\sim \mc{O}\left(\vo^{-p}\right) \delta \phi+ \mc{O}(1)\, \delta \chi_{\vo}+\mc{O}\left( \sqrt{\vo}\right) \delta \chi_s\,,
\label{eq:canNorm3}
\ee
where $\phi$ is the quintessence field, while $\chi_{\vo}$ and $\chi_s$ are two heavy fields corresponding respectively to the volume
mode and the small rigid divisor $\tau_s$. The exact coefficients of (\ref{eq:canNorm3}) will be computed in appendix \ref{AppendixA}
at leading order in an inverse volume expansion.
What is crucial, in the previous expression, is that the canonically normalised quintessence field $\phi$ provides
only a suppressed contribution to $\delta \tau_s$, essentially determined by the ratio of the leading and subleading scalar potentials of
eqs (\ref{Vlead}) and (\ref{Vsub}) (recall also the general discussion of section \ref{subscoupl}). Since $\tau_s$ controls
the properties of SM physics, this implies that the coupling of the quintessence field to Standard Model particles is very suppressed.

Before passing to prove this fact explicitly, let us discuss a technical issue:  during the quintessence dynamics
the two heavy fields $\chi_{\vo}$ and $\chi_s$ sit at their minima,  but the rolling scalar $\phi$ can be far away from its minimum.
Hence the fibre modulus has to be canonically normalised not just close to its minimum but for an arbitrary point in moduli space.
This is a complicated computation whose result has already been qualitatively discussed in \cite{Cicoli:2012cy}
where the authors used the poly-instanton effects for deriving a model of modulated reheating.
The final upshot is that in regions far away from the minimum for large values of $\tau_f$,
the poly-instantons get more suppressed, and so the mixing with the other moduli also gets more suppressed.
Thus the coupling of the fibre modulus to ordinary matter gets weaker in regions far from the minimum.
In fact, if $\tau_f$ is far from its minimum then:
\be
e^{-2\pi \tau_f} = e^{-2\pi \langle\tau_f\rangle} e^{-2\pi \hat\tau_f} \sim \vo^{-p} e^{-2\pi \hat\tau_f}\,,
\ee
and so the mixing term in (\ref{eq:canNorm3}) would be even more suppressed than $\vo^{-p}$.

\subsection{Coupling to gauge bosons}

We start by estimating the couplings to gauge bosons following \cite{ADDstrings}.
Recalling that $\tau_s$ is the Standard Model cycle, the gauge kinetic terms are given by:
\be
\mc{L}_{\rm gauge}=-\frac{\tau_s}{M_P}F_{\mu\nu}F^{\mu\nu}\,.
\ee
After expanding $\tau_s$ around the minimum ($\tau_s=\langle\tau_s\rangle+\delta \tau_s$)
and performing the canonical normalisation of the field strength tensor, we find that the interaction term is:
\be
\mc{L} \supset \frac{\delta \tau_s}{4 M_P \langle\tau_s\rangle} G_{\mu\nu}G^{\mu\nu}.
\ee
From eq. (\ref{eq:canNorm3}) it follows that:
\be
\frac{\delta \tau_s}{\langle\tau_s\rangle}
\sim \mc{O}\left(\vo^{-p}\right) \delta \phi\,,
\ee
and so we find:
\be
\mc{L} \supset \mc{O}\left(\frac{1}{M_P\,\vo^p}\right)\delta\phi \,G_{\mu\nu}G^{\mu\nu}.
\ee
We conclude that the coupling between the fibre modulus and the SM gauge bosons is weaker-than-Planckian for $p\sim \mc{O}(1)$,
in accordance with \cite{ADDstrings}, and therefore poses no issues with the constraints from fifth-force experiments.

\subsection{Coupling to fermions}

Let us now turn our attention to the coupling between the quintessence field and the Standard Model fermions, which we denote as $C_i$.
The relevant part of the SUGRA Lagrangian for the computation of this coupling is:
\be
\mc{L} \supset \tilde{K}_{ij}\overline{C_i}i\gamma^\mu \partial_\mu C_j + \tilde{K}_{H\overline{H}} \partial^\mu H \partial_\mu \overline{H}+ e^{K/2}\lambda_{ij}C_i C_j H,
\ee
where $\tilde{K}_{ij}$ and $\tilde{K}_{H\overline{H}}$ are generic functions of the moduli fields. Very little can be said about this theory without specifying $\tilde{K}_{ij}$ and $\tilde{K}_{H\overline{H}}$, and so we follow the work of \cite{Conlon:2006tj}, where the leading modular dependence of these two functions has been estimated through simple scaling arguments. We simplify the set-up further by considering diagonal K\"ahler metrics and Yukawa couplings:
\be
\mc{L} \supset \tilde{K}_{ii}\overline{C_i}i\gamma^\mu \partial_\mu C_i + \tilde{K}_{H\overline{H}} \partial^\mu H \partial_\mu \overline{H}+ e^{K/2}\lambda \overline{C_i} C_i H.
\ee
Following \cite{Conlon:2006tj} we take:
\be
\tilde{K}_{ii}\sim \tilde{K}_{H\overline{H}}\sim \frac{\tau_s^{1/3}}{\mc{V}^{2/3}}.
\ee
Expanding $\tau_s$ and $\mc{V}$ around their VEVs as $\tau_s=\langle\tau_s\rangle+\delta\tau_s$ and $\mc{V}=\langle\mc{V}\rangle+\delta\mc{V}$
we can write the K\"ahler metrics as:
\be
\tilde{K}_{ii}\sim \tilde{K}_{H\overline{H}}\sim\tilde{K}_0\left(1+\frac{\delta\tau_s}{3\langle\tau_s\rangle}
-\frac{2\delta\vo}{3\langle\vo\rangle}\right) \qquad \text{where} \qquad \tilde{K}_0\equiv \frac{\langle\tau_s\rangle^{1/3}}{\langle\vo\rangle^{2/3}}.
\ee
This allows us to define the canonically normalised fields $c_i=\sqrt{\tilde{K}_0} C_i$ and $h=\sqrt{\tilde{K}_0} H$ and rewrite the Lagrangian as:
\be
\mc{L} \supset \left(1+\frac{\delta\tau_s}{3\langle\tau_s\rangle}-\frac{2\delta\vo}{3\langle\vo\rangle}\right) \overline{c_i}i\gamma^\mu \partial_\mu c_i + \partial^\mu h \partial_\mu \overline{h}+ \frac{\lambda}{\mc{V} \tilde{K}_0^{3/2}} \overline{c_i} c_i h.
\ee
Setting $h=\langle h\rangle$, expanding $1/\mc{V}\sim 1/\langle\mc{V}\rangle(1-\delta \vo/ \langle\vo\rangle)$ and defining the fermionic mass as $m_c\equiv \frac{\lambda \langle h \rangle}{\langle \tau_s\rangle^{1/2}\tilde{K}_0^{3/2}}$ this becomes:
\be
\mc{L} \supset \overline{c_i}(i\gamma^\mu \partial_\mu +m_c)c_i +\left(\frac{\delta\tau_s}{3\langle\tau_s\rangle}-\frac{2\delta\vo}{3\langle\vo\rangle}\right)\overline{c_i}(i\gamma^\mu \partial_\mu +m_c)c_i -m_c \overline{c_i} c_i \left( \frac{\delta \tau_s}{3\langle\tau_s\rangle} +\frac{\delta \vo}{3\langle\vo\rangle} \right).
\ee
Note that the first term is the standard Lagrangian for a free fermion of mass $m_c$, the second term vanishes once we impose the equations of motion and the third term encodes the interaction between moduli and Standard Model fermions. Expressing the moduli in terms of the canonically normalised fields near the minimum, we find that the dominant term in the interaction with the quintessence field comes from the expression (\ref{eq:canNorm3}) for $\delta \tau_s$ term, since to leading order $\delta \vo$  does not depend on the canonically normalised fibre modulus. This implies that the leading contribution to the dimensionless coupling between ordinary fermions and the quintessence field is:
\be
\lambda_{\phi \overline{c}c}\sim \mc{O}\left(\frac{m_c}{M_P \vo^p}\right),
\ee
which is weaker-than-Planckian for $p\sim \mc{O}(1)$ like the coupling to Standard Model gauge bosons.

The fact that the couplings to gauge bosons and fermions are suppressed by a scale higher than the Planck mass means that the current model is able to comfortably accommodate the constraints from fifth-force experiments, see e.g. \cite{Adelberger:2003zx}, and provide a controlled string theoretical description of quintessence.

\section{Dynamics of the quintessence-dark matter system}
\label{Dynamics}

In this section we study the dynamics of the quintessence-dark matter system by following the autonomous phase plane analysis originally developed in \cite{Copeland:1997et}. We start by analysing a model in which we neglect any  direct coupling  between the
 two dominant components of the energy density (this
since their direct coupling is weaker-than-Planckian).
Then we look at the case where there is a direct,  Planck strength coupling between dark matter and dark energy,
while at the same time
dark energy couples extremely  weakly to ordinary matter. In this way, we will outline a proposal of
how to realise a coupled quintessence scenario
within string theory.

We consider a flat FRW Universe filled with quintessence as described by the fibre modulus and a pressureless dark matter component. The time evolution of the system is determined by the Raychaudhuri equation, the continuity equation for the dark matter component, the Klein-Gordon equation for a homogeneous scalar field and the Friedmann constraint:
\bea
\dot{H}=-\frac{1}{2M_P^2}\left( \rho_{DM}+\dot{\phi}^2\right),\\
\dot{\rho}_{DM}=-3 H \rho_{DM}+Q,\label{eq:DMcons}\\
\ddot{\phi}+3 H \dot{\phi}+\frac{d V}{d \phi}=\frac{Q}{\dot{\phi}}\\
\Omega_{DM}+\Omega_{DE}=1,
\eea
where $\Omega\equiv \frac{\rho}{3 H^2 M_P^2}$ and  $Q$ parameterises the direct interaction between the dark matter and the quintessence components.
Using $N=\ln a$ as the time variable and defining:
\be
x \equiv \frac{ \phi'}{M_P\sqrt{6}},\qquad y^2\equiv \frac{ V}{3 M_P^2 H^2},
\label{eq:xyDefn}
\ee
the equations of motion can be cast in the following form:
\bea
x'(N)=-3x-\frac{\partial V/\partial \phi}{V} \sqrt{\frac{3}{2}}y^2+\frac{3}{2}x(2 x^2+(1-x^2-y^2)) + \frac{Q}{6 H^3 M_P^2 x},\label{eq:x}\\
y'(N)=\frac{\partial V/\partial \phi}{V} \sqrt{\frac{3}{2}}x y +\frac{3}{2}y (2 x^2 + (1-x^2-y^2)),\\
H'(N)=-\frac{3 H}{2}(2 x^2 +(1-x^2-y^2)),\\
\phi'(N)=\sqrt{6} x.
\eea
In what follows we will both search analytically for instantaneous fixed points of the system and study the phase space trajectories numerically.

\subsection{Decoupled quintessence and dark matter}
\label{Decoupled}

We first analyse the case where a direct coupling between dark matter and dark energy is negligible, which corresponds to setting $Q=0$. The system possesses two non trivial instantaneous fixed points, $x'(N)=y'(N)=0$, whose properties are listed in table \ref{tab:fixedPoints}. These correspond to motion in the two distinct regions of the potential: point $\mc{A}$ corresponds to motion in the poly-instanton dominated part of the potential (labelled II in figure \ref{fig:Potential}) whereas
point $\mc{B}$ describes motion in the region where the loop potential dominates (labelled region III in figure \ref{fig:Potential}).
Note that the flatness of the poly-instanton potential implies that $\delta\equiv V'_{\rm poly}/V_{\rm poly}\sim 0$,
and so the behavior of the fibre modulus approaches that of a pure cosmological constant.

\begin{table}[h]
\begin{center}
\begin{tabular}{c|c|c|c|c}
Point & ($x_c,y_c$) & Stability & $\Omega_\phi$ & $\omega_\phi$\\
\hline
$\mc{A}$& $(\delta,\sqrt{1-\delta^2})$& stable node&1&$-1$\\
$\mc{B}$& $(2/\sqrt{18},\sqrt{14/18})$& stable node&1&$-5/9$\\
\hline
\end{tabular}
\end{center}
\caption{Fixed points for the quintessence dark-matter system.}
\label{tab:fixedPoints}
\end{table}

\begin{figure}[h!]
	\centering
	\begin{minipage}[b]{0.45\linewidth}
	\centering
\includegraphics[width=1\textwidth]{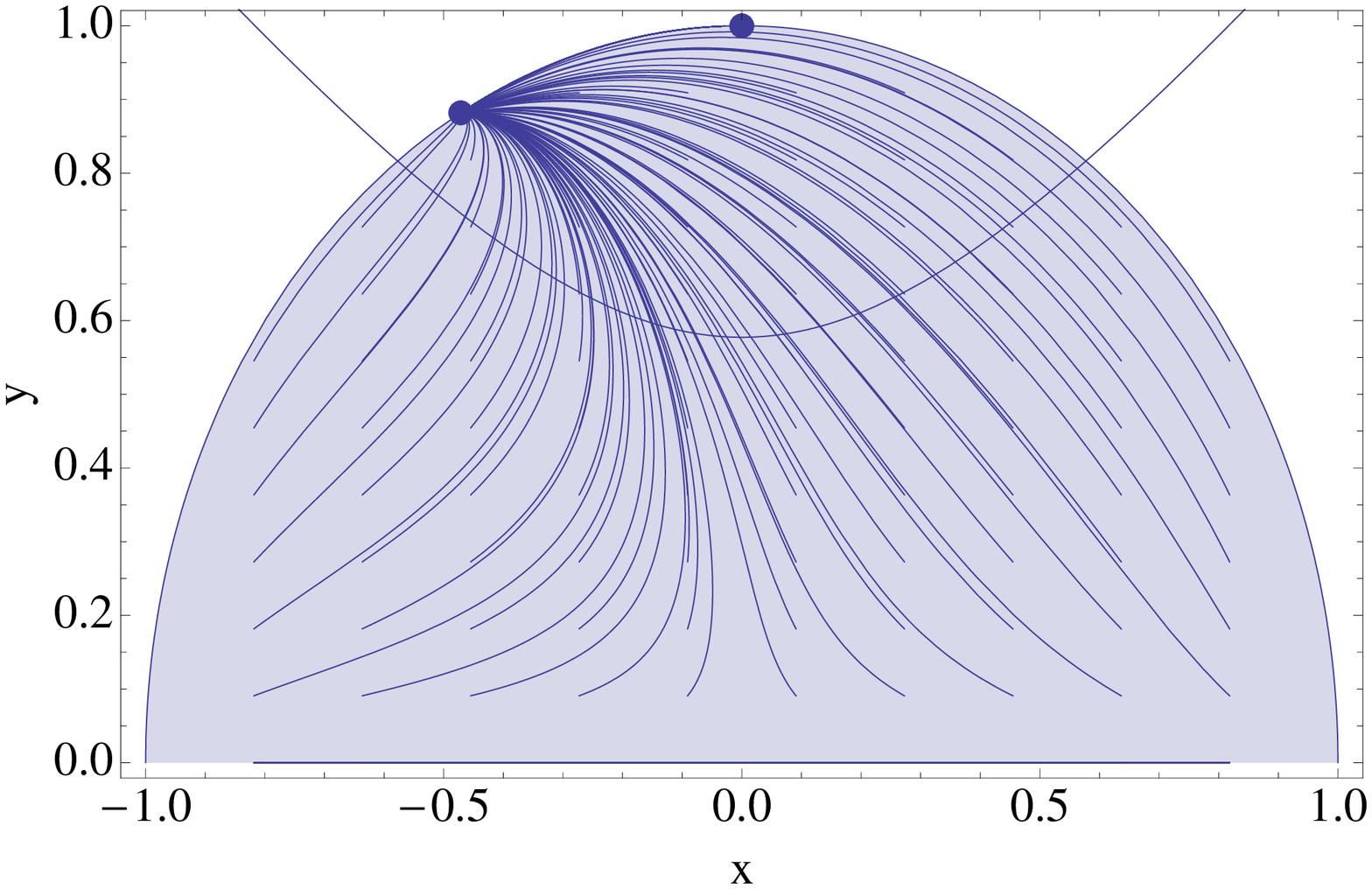}
    \end{minipage}
	\hspace{0.05cm}
	\begin{minipage}[b]{0.45\linewidth}
	\centering
\includegraphics[width=1\textwidth]{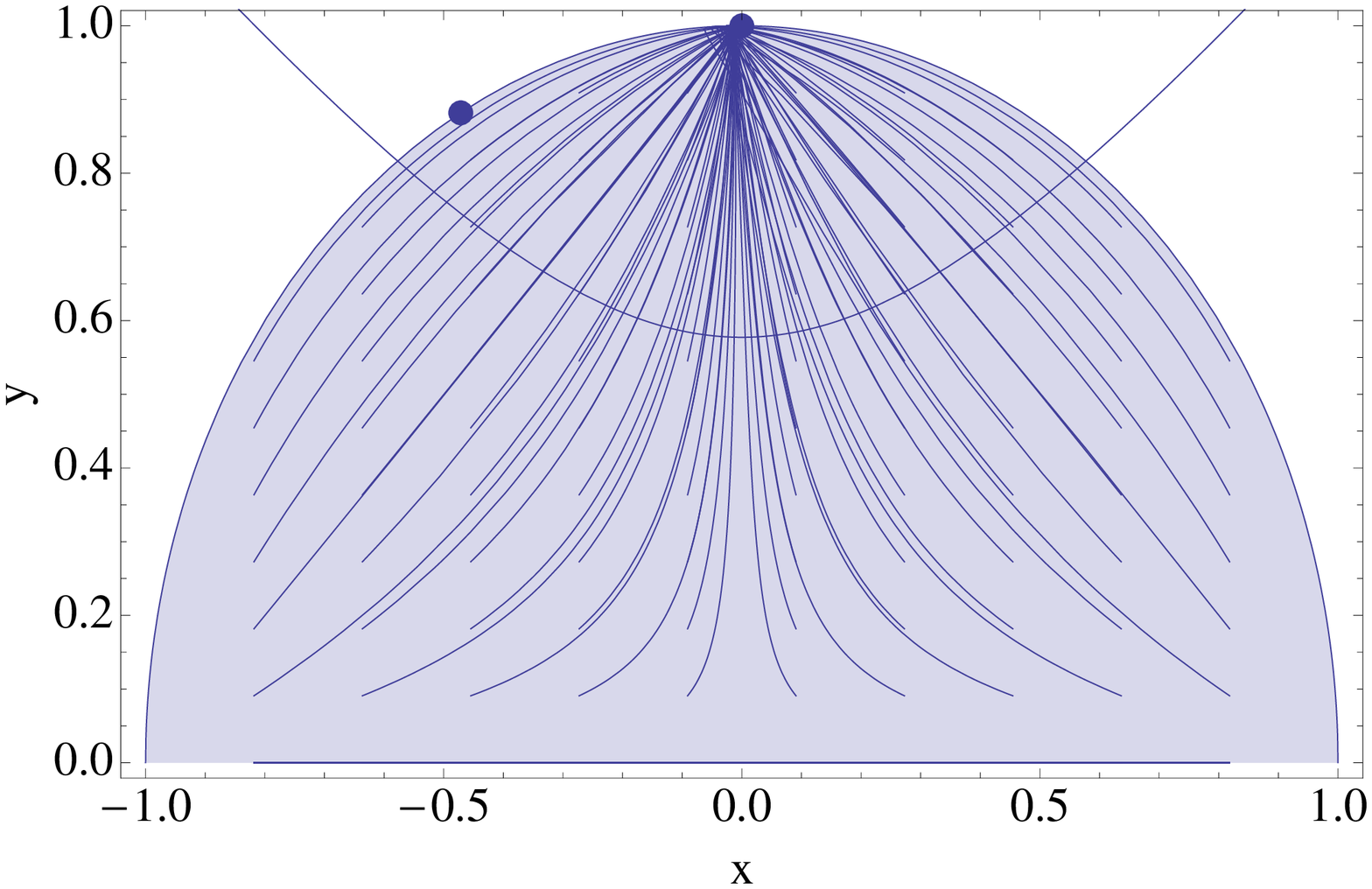}
	\end{minipage}
	\hspace{0.05cm}
	\caption{Phase plane evolution of the quintessence dark-matter system. Left: the fibre modulus starts in the loop dominated region of the potential;
right: the fibre modulus starts in the poly-instanton dominated region of the potential.}
	\label{fig:phasePlane}
\end{figure}

From the definitions of eq. (\ref{eq:xyDefn}), one sees that $x(N)$ ($y(N)$) is the ratio of the scalar field's kinetic (potential) energy to the total energy.
We illustrate the behaviour of the system in figure \ref{fig:phasePlane} for different sets of initial conditions. The region above the parabola corresponds to accelerating solutions and the dots represent the two instantaneous fixed points described in table \ref{tab:fixedPoints}.

We see that if the fibre modulus starts its evolution in the part of the potential that is dominated by the loop effects,
the system will evolve towards the fixed point $\mc{B}$. This fixed point corresponds to an accelerated expansion solution, where the energy density is dominated by the scalar field, as preferred by current cosmological data. However the equation of state parameter, which is generically given by:
\be
\omega_\phi=\frac{x-y}{x+y},
\ee
is different from 1 for the fixed point $\mc{B}$. Thus, given the current observational constraints, this fixed point is not a viable candidate for dark energy.  As the fibre modulus rolls down the loop generated potential it eventually enters a region in field space where the loop potential is suppressed relative to the poly-instanton generated potential. In this region the relevant fixed point to which the system will  converge is $\mc{A}$, which corresponds to slow roll dynamics in the poly-instanton dominated part of the potential. The fixed point $\mc{A}$ describes a flat Universe filled with a negative pressure fluid with $\omega=-1$.  Even though such trajectories display the same asymptotic behaviour in the future as the observed Universe, a numerical analysis of the system suggests they never go through a phase in which the conditions $\Omega_{DE}\sim 0.7$ and $\omega_\phi\sim-1$ are simultaneously verified. It then follows that trajectories that start with the quintessence field deep in the loop dominated region of potential do not provide a viable description of the late time Universe. On the other hand, if the initial conditions are such that the fibre modulus starts its evolution in a region of the potential where the loop effects are negligible, it will evolve towards the fixed point $\mc{A}$ without going through $\mc{B}$ first.  As can be seen in figure \ref{fig:phasePlane}, these trajectories can go through the region of phase space where the energy density and the equation of state parameter
of the quintessence field are in agreement with observational data. We then conclude that there are phase space trajectories for which the system evolves to a state where it provides a viable description of an accelerated expanding Universe, namely the ones corresponding to initial field values in the poly-instanton plateau. It is worth noting that from the discussion in section \ref{QuintPot} the width of the poly-instanton plateau is determined by the ratio $V_{\rm poly}/V_{\rm loop}\sim \vo^{1-p} $ which can easily be $\gg 1$ given the large volumes required for quintessence phenomenology, $\vo\sim 10^{30}$. There is therefore an appreciable range of initial conditions for the fibre modulus that yield a viable quintessential behaviour.

When looking at the late time composition of the Universe one is confronted with the coincidence problem: why is the dark energy density comparable to the dark matter density today? Dynamical models of dark energy often address this issue by allowing for scaling solutions: fixed points where the ratio of dark energy to dark matter density is constant. The existence of these scaling solutions is determined by the steepness of the dark-energy potential ($V'/V$) and by the equation of state for the matter component. In a Universe with pressureless cold dark matter and the fibre modulus playing the r\^ole of quintessence, there are no such fixed points as both regions of the quintessence potential are too shallow to support them. Therefore we are led to the conclusion that even though our model provides a realistic description of the current state of the Universe, it does not allow for a dynamical explanation of the coincidence problem: we just happen to be around at a time when $\Omega_{DE}$ and $\Omega_{DM}$ are of the same order and the Universe will eventually evolve to a state where it is filled with quintessence.

In these considerations about the distant future evolution of the system we have so far ignored the presence of the minimum of the potential,
region I in fig. \ref{fig:Potential},  to the left of the poly-instanton plateau.
Its presence implies that in the very distant future the fibre modulus will eventually fall into its minimum, after which will follow a phase of oscillatory dynamics, in a process similar to post-inflationary reheating, albeit at a much lower energy scales.
Since, ultimately, acceleration ends with the scalar sitting at its Minkowski minimum, we avoid the
problems raised in \cite{Hellerman,Fischler} with respect of embedding asymptotic de Sitter space in string theory.

\subsection{Coupled quintessence and dark matter}
\label{Coupled}

We now analyse how a coupled quintessence-dark matter system affects the dynamics of the Universe \cite{Amendola}.
Possible Planck suppressed couplings between dark energy and dark matter are allowed in our set-up
if dark matter is constituted by bulk KK modes or moduli fields.

Let us start discussing possible dark matter candidates:
phenomenological constraints on dark matter from collider physics, astrophysics and cosmology  in this  class of scenarios
have been discussed in \cite{ADDstrings}. Here we briefly summarise  the main conclusions.
The requirement of avoiding BBN constraints, associated with the fact that the SM brane cools too quickly
due to evaporation of energy into the bulk, sets a severe bound on the reheating temperature: $T_{\rm RH}\lesssim 100$ MeV.
If this constraint is satisfied, there are no issues with BBN, but the relic KK modes, if stable, can still overclose the Universe.
In order to avoid this, the previous bound becomes more severe, bringing the reheating temperature at most around a few MeV,
just above the BBN temperature. On the other hand, if the KK modes are not stable and decay into photons,
then one can have problems with possible distortions of the diffuse MeV $\gamma$-ray background.

Notice that these bounds on the reheating temperature imply that we cannot have a thermal bath up to the weak scale,
and so the standard thermal WIMP picture for dark matter does not work.
In addition, we do not have neutralini in the 4D low-energy spectrum since supersymmetry is badly broken on the SM brane.
In our case we have two kinds of KK modes: 6D KK modes and 10D KK modes. The 10D KK modes decay whereas the 6D KK modes are so long lived that can be considered as almost stable with respect to the age of the Universe. Hence given that the 6D KK modes are almost stable, we need to have a reheating temperature just above the MeV scale. The advantage is that the 10D KK modes decay predominantly to 6D KK modes,
and so their decay to photons is suppressed. Thus we have no problems associated with $\gamma$-ray background distortion.

Higher 6D KK states can be produced from the thermal bath but they quickly decay to lower 6D KK states, which are then long-lived. There are also two light moduli which are stable with respect to the age of the Universe. One is the volume mode which we can suppose to be the inflaton, so that inflation takes place when the gravity scale is high, whereas at the end of inflation, the string scale is around the TeV. This implies that the volume mode is very light with a mass of the order meV. Hence this modulus can be produced in the thermal bath (or by simple scalar oscillation) and then suffers from the cosmological moduli problem (CMP). The other light modulus is the fibre divisor which drives quintessence.

Therefore, we conclude that in our scenario dark matter can be realised in terms of a mix of bulk 6D KK modes and the volume modulus, once the CMP
associated with the latter is solved (for example via a second low-energy inflationary period as in thermal inflation).
The quintessence field would then indeed couple to dark matter, for example 6D KK modes, with Planck suppressed trilinear metric couplings of the form:
\be
\frac{\psi}{ M_P}   \partial \phi \partial \phi\,,
\ee
where $\phi$ is the quintessence field and $\psi$ the 6D KK mode.

After this qualitative discussion to motivate the coupling of dark energy to dark matter in our set-up, we can write the dark matter energy density as a function of the fibre modulus as:
\be
\rho_{DM}= \rho_0 f(\phi),
\ee
where $\rho_0$ obeys the normal conservation equation:
\be
\dot{\rho}_0=-3 H \rho_0,
\ee
and $f(\phi)$ parameterises the interaction between dark matter and the quintessence field. The conservation equation for dark matter is then given by:
\be
 \dot{\rho}_{DM}=-3 H \rho_{DM}+\frac{\partial f}{\partial \phi}\frac{\dot{\phi}}{f(\phi)}\rho_{DM}.
\label{eq:DMconsInt}
 \ee
By comparing eqs. (\ref{eq:DMcons}) and (\ref{eq:DMconsInt}) we define $Q\equiv-\frac{\partial f}{\partial \phi}\frac{\dot{\phi}}{f(\phi)}\rho_{DM}$. In order to look for fixed points analytically, one must be able to write $\frac{\partial f}{f(\phi)\partial \phi}$ in terms of $x(N), y(N)$ and other dimensionless parameters.
One simple case might be:
\be
f(\phi)= e^{ \alpha \phi/M_P}\,,
\ee
since this form for the coupling function implies that:
 \be
\frac{\partial f}{\partial \phi}\frac{\dot{\phi}}{f(\phi)}\rho_{DM}=\frac{\alpha}{M_P} \dot{\phi}\rho_{DM}\,,
 \ee
allowing for an analytical study of the phase plane dynamics of the system. This is exactly what one expects to find if the dark matter mass is a function of the fibre modulus $\tau_f$,
\be
m_{DM}\sim \tau_f ^m \sim e^{2 m \phi/\sqrt{3}}
\ee
where in the last step we have written $\tau_f$ in terms of the canonically normalised field $\phi$.

It then follows that $Q=-\alpha \sqrt{6} x H \rho_{DM}$ and eq. (\ref{eq:x}) becomes:
 \be
 x'(N)=-3x-\frac{\partial V/\partial \phi}{V} \sqrt{\frac{3}{2}}y^2+\frac{3}{2}x(2 x^2+(1-x^2-y^2)) -\alpha \sqrt{3/2}(1-x^2-y^2).
\ee

The phase space for models with this form of the direct coupling Q has been studied in detail in \cite{CoupledDEDM}, where a phenomenologically interesting fixed point located at $(x_c,y_c)=(\frac{\sqrt{6}}{2 b},\frac{\sqrt{9+6 \alpha b}}{\sqrt{6}b})$ was found. Note that we have defined $\lambda\equiv -M_P\frac{\partial V} {V\partial \phi} $ and $b\equiv \lambda +\alpha$. Following \cite{CoupledDEDM} we find that this fixed point exists if:
\be
b^2\ge\frac{3}{2}\qquad\text{and}\qquad -\frac{3}{2} \le \alpha b \le b^2-3
\label{eq:existence}
\ee
and is stable if:
\be
b^2\ge\frac{3}{2}\qquad\text{and}\qquad \left(-\frac{3}{2}< \alpha b < -B_{sgn(b)}\qquad\text{or}\qquad  -B_{-sgn(b)}< \alpha b < b^2-3 \right),
\label{eq:stability}
\ee
where:
\be
B_{\pm}=\frac{2 b[\pm (b^2-3/2)^{3/2}-b(b^2-39/8)]}{4b^2+3/4}.
\ee

As in the decoupled case, we proceed by considering two different regimes, depending on which term in the potential dominates.
For each case we shall then investigate whether the above conditions are met. In the regime in which the poly-instanton effects dominate over the loop potential (region II of figure \ref{fig:Potential}), $\lambda \sim 0$ to a very good accuracy. One sees that it is impossible to satisfy the existence conditions given by eq. (\ref{eq:existence}). We then conclude that the poly-instanton potential is unable to support an interacting tracking solution. However if  one looks at the regime where the loop generated potential dominates (region III of figure \ref{fig:Potential}), one has $\lambda =- 2/\sqrt{3}$ and one sees that there are values of $\alpha$ for which a stable accelerating tracking solution exists. This region however is not very wide due to the requirement of stability. In fact we find that it is roughly $-\frac{5}{2\sqrt{3}}-0.55 <\alpha<-\frac{5}{2\sqrt{3}}$.  We must note that this is not so much a consequence of the shape of the potential as of the explicit form of $Q$ we are assuming. Furthermore, we see that in this narrow stability region  $0.93<\Omega_{DE}<1$, where the lower bound corresponds to $\alpha= -\frac{5}{2\sqrt{3}}-0.55 $ and the upper bound to  $\alpha= -\frac{5}{2\sqrt{3}}$.  It is worth noting that for $\alpha <0$ the dark matter mass goes like $m_{DM}\propto 1/\tau_f^m$.

\begin{figure}[h!]
	\centering
	\begin{minipage}[b]{0.45\linewidth}
	\centering
\includegraphics[width=1\textwidth]{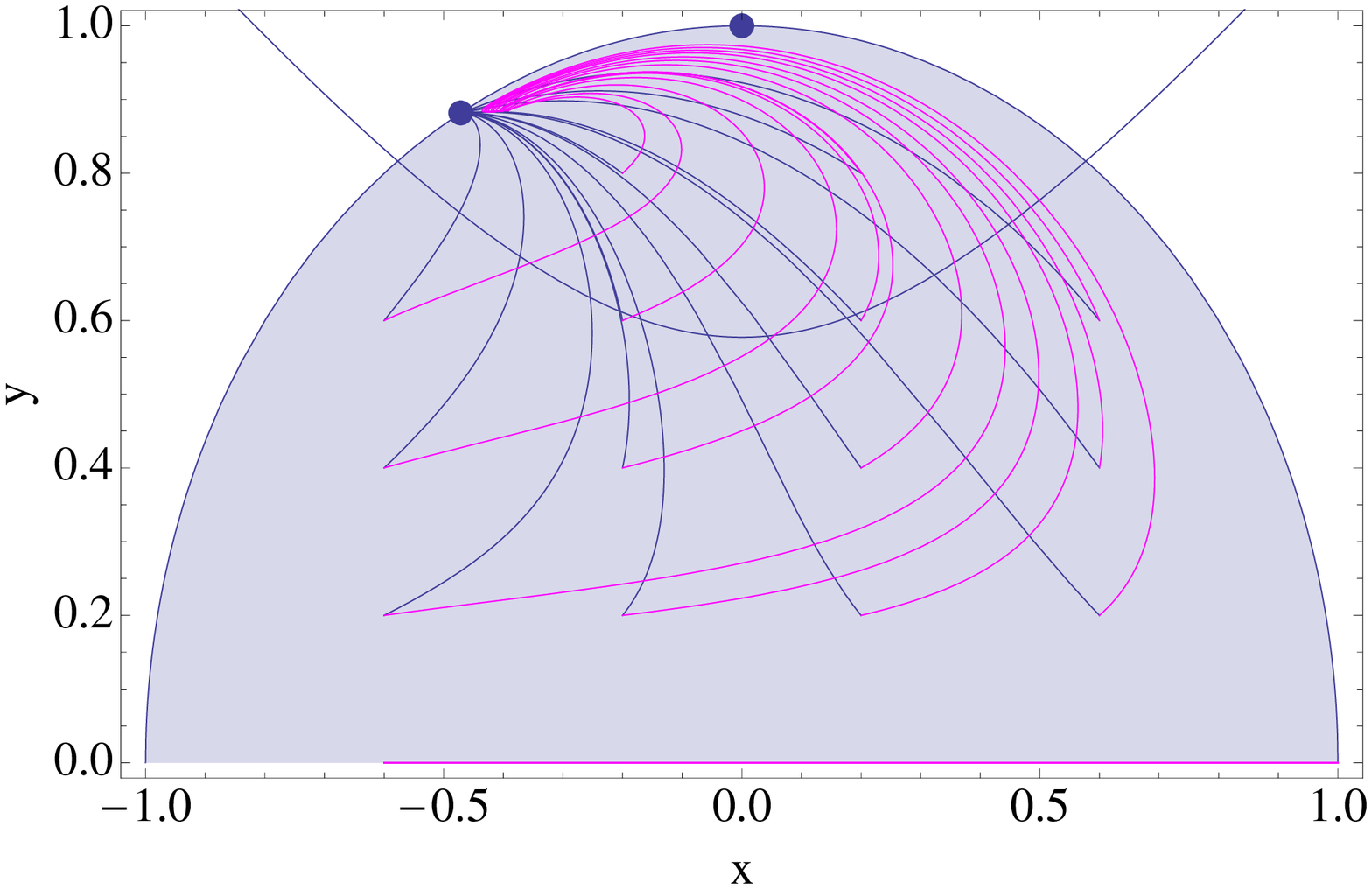}
    \end{minipage}
	\hspace{0.05cm}
	\begin{minipage}[b]{0.45\linewidth}
	\centering
\includegraphics[width=1\textwidth]{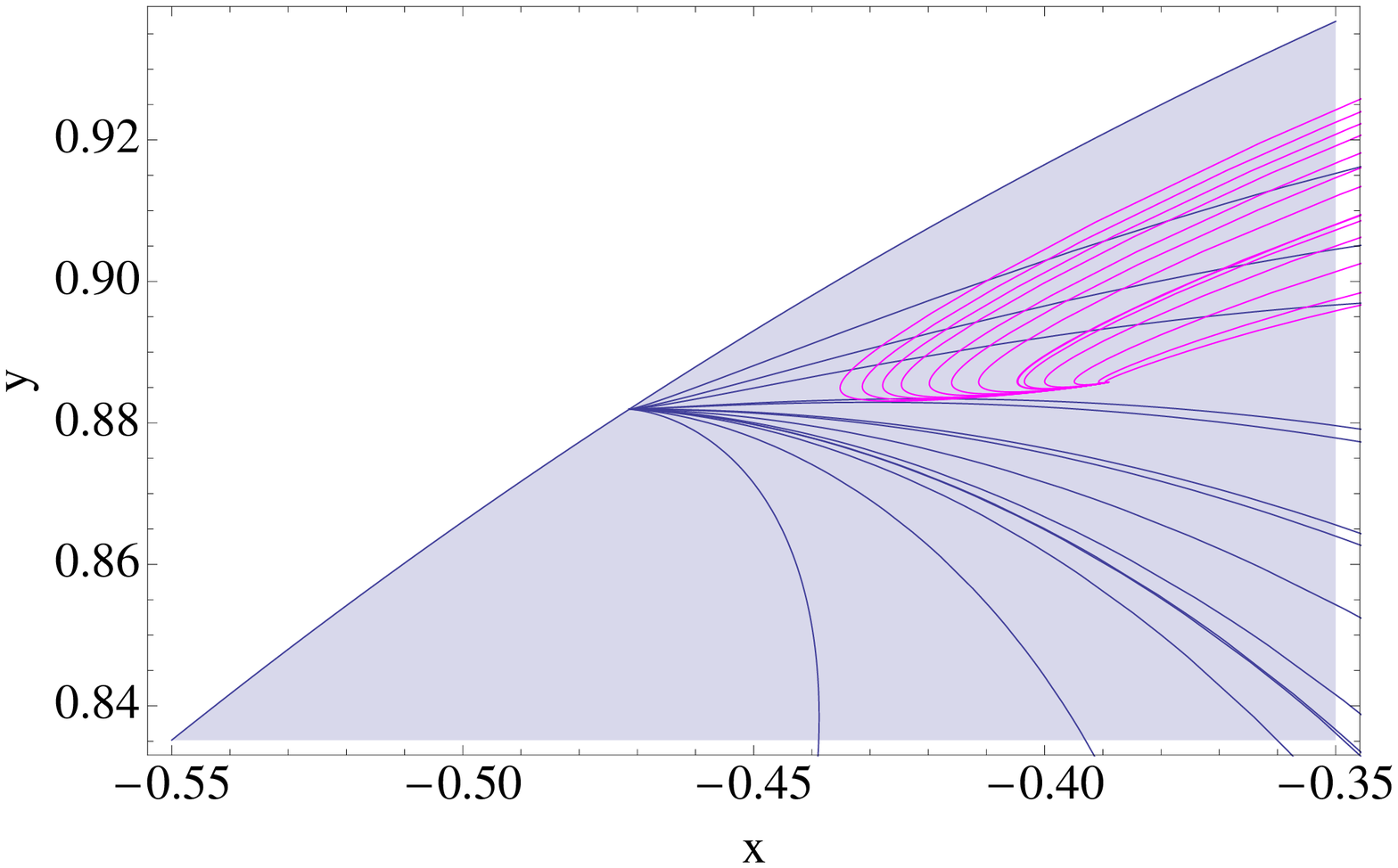}
	\end{minipage}
	\hspace{0.05cm}
	\caption{Phase plane evolution of the quintessence dark-matter system with interaction and $F_{\rm poly}\gg F_{\rm loop}$}
	\label{fig:phasePlaneLoopInt}
\end{figure}

The behavior described above is illustrated in fig. \ref{fig:phasePlaneLoopInt}, where it is clear that the presence of the interaction dramatically changes the trajectories in phase space. One also sees that the coupled system (magenta) converges to a fixed point which is different from the one of the decoupled case (blue lines).

Once one considers the full potential, the fixed point described above is transient since as the field rolls down the loop generated potential it eventually enters the poly-instanton dominated region. So the phase-space trajectories  start by converging to the fixed point $\mc{B}$  before evolving to the cosmological constant fixed point $\mc{A}$.

\begin{figure}[h!]
	\centering
	\begin{minipage}[b]{0.45\linewidth}
	\centering
\includegraphics[width=1\textwidth]{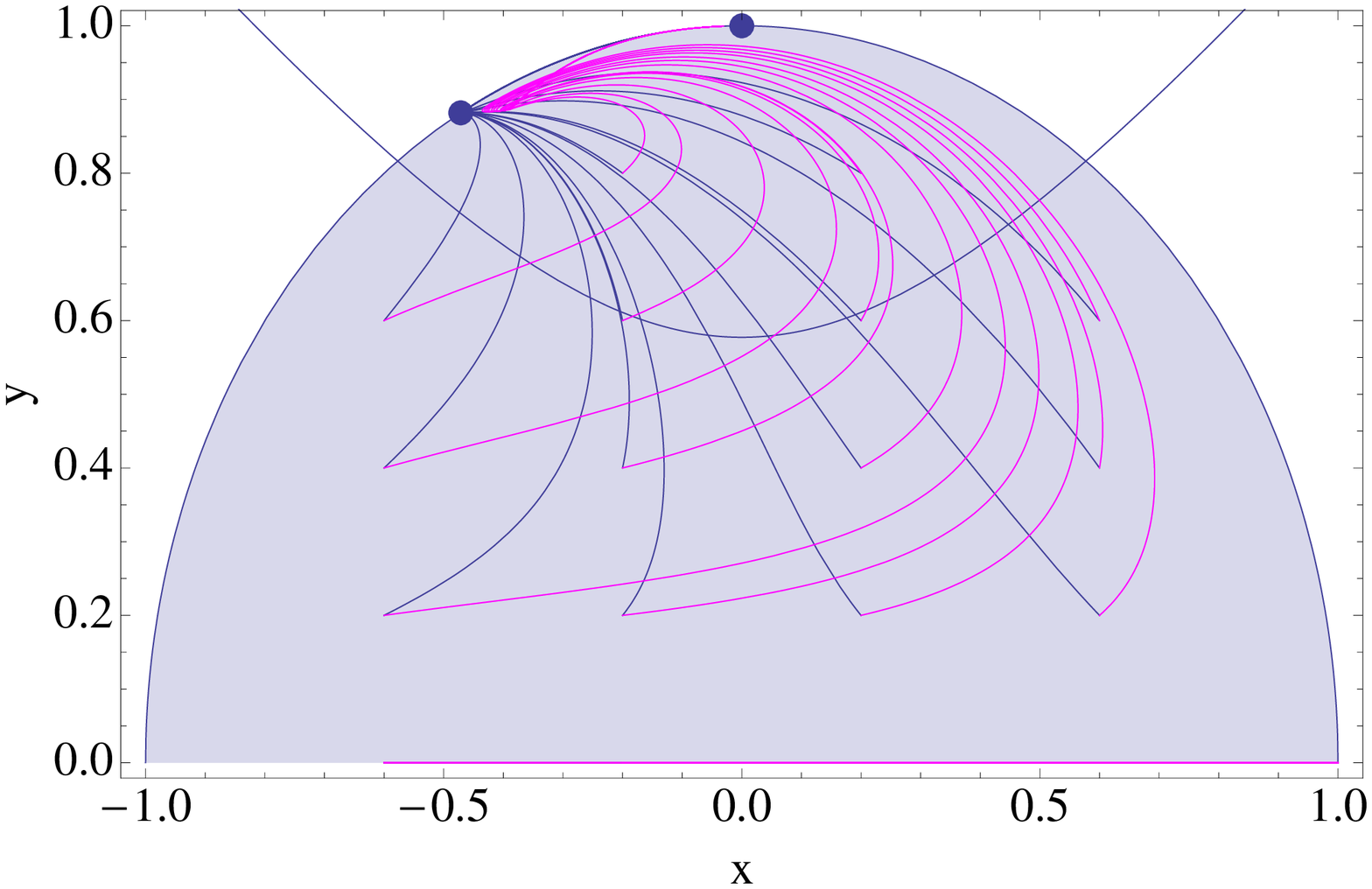}
    \end{minipage}
	\hspace{0.05cm}
	\begin{minipage}[b]{0.45\linewidth}
	\centering
\includegraphics[width=1\textwidth]{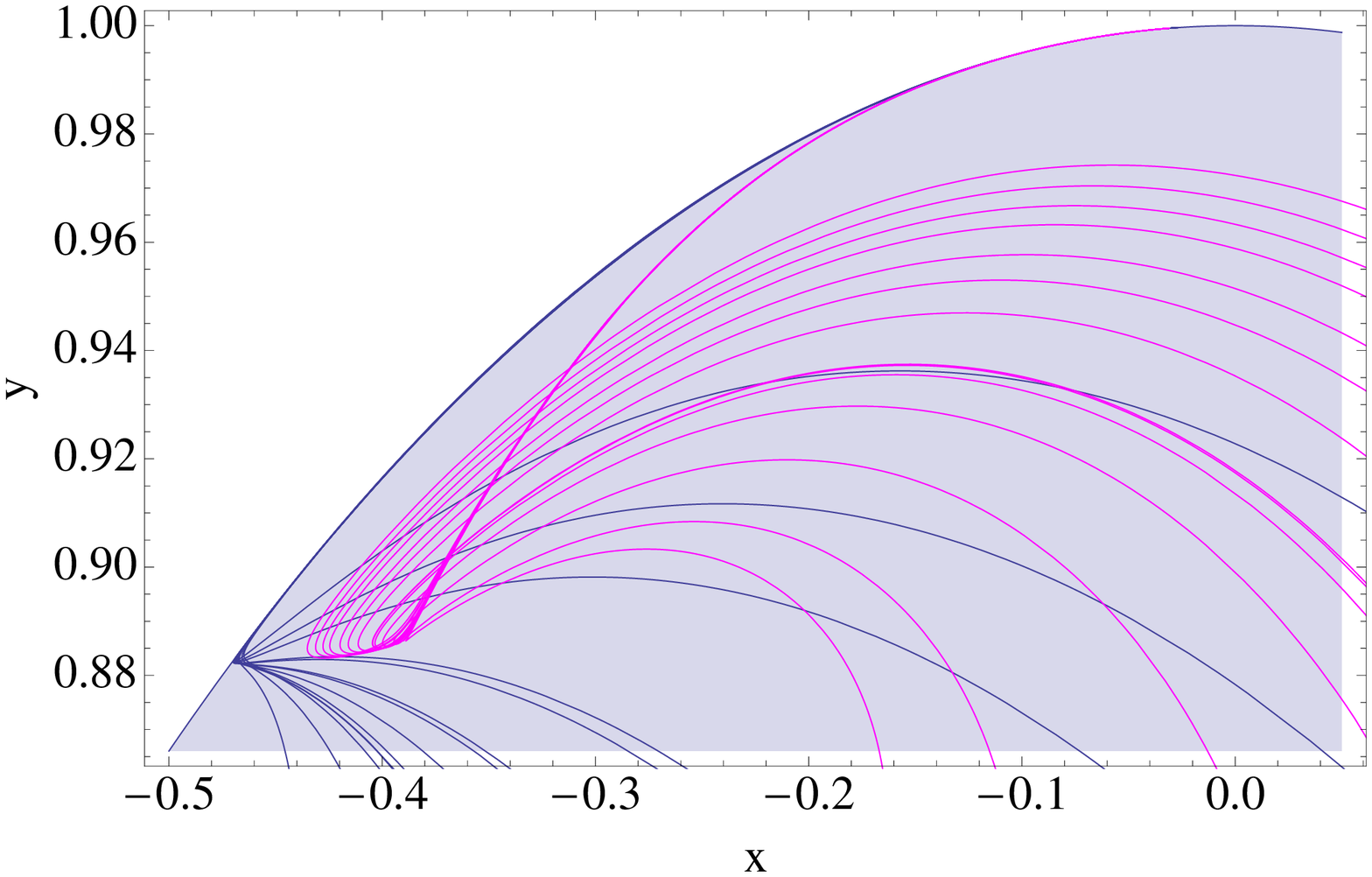}
	\end{minipage}
	\hspace{0.05cm}
	\caption{Phase plane evolution of the quintessence dark-matter system with interaction and $F_{\rm poly} \sim 10^3 F_{\rm loop}$}
	\label{fig:phasePlaneFullInt}
\end{figure}

As in the non interacting case, the fibre modulus is able to provide suitable description of dark energy even though it does not address the coincidence problem. The introduction of the non minimal interaction term extends the existence of viable trajectories for cases when the field starts its evolution deep in the loop dominated part of the potential.

\section{Conclusions}
\label{Conclusions}

In this paper, we presented a new quintessence model embedded in string theory
where the small mass of the rolling scalar field is technically natural and its
coupling to Standard Model degrees of freedom is naturally weaker-than-Planckian,
thus avoiding fifth-force constraints.

We focused on a type IIB LARGE Volume Scenario with a very anisotropic shape of the extra dimensions
since two of them are stabilised at values exponentially larger than the other four \cite{ADDstrings}. This leads to
a string derivation of SLED scenarios \cite{SLED} which have already turned out to be very promising
for obtaining models of quintessence \cite{Albrecht:2001xt}. In these scenarios,
if TeV-scale gravity provides the solution to the hierarchy problem,\footnote{Notice that all the scales of our model
are set by the value of the overall volume $\vo$ which has an exponential dependence on the underlying parameters,
and so it can be easily set to arbitrarily large values without tunings.}
the gravitino mass $m_{3/2}$ turns out to be of the observed order of magnitude of the cosmological constant:
$m_{3/2}\sim\Lambda$. Therefore,
the current value of dark energy $\Lambda$ can be naturally reproduced if the quintessence field
develops a potential at order $V \sim m_{3/2}^4$.

This mechanism, which ties the solution of the hierarchy problem to the one of the cosmological constant problem, 
relies on the existence of a compensation between the SM brane tension and its back-reaction. A possibility to achieve this could
be to directly apply to our set-up the ideas developed in 6D SLED constructions.  This, in turn, depends on the details of the 6D effective supergravity theory for our
anisotropic compactifications; on the other hand, this intermediate theory has not yet been explicitly identified in \cite{ADDstrings}.
We leave a proper  understanding of these interesting open issues for study in the near future.

In this paper, we showed that the fibre divisor $\tau_f$ indeed develops a potential at order $V \sim m_{3/2}^4$
via tiny poly-instanton corrections to the superpotential and closed string loop contributions to the
K\"ahler potential. Furthermore, this field is radiatively stable since the bulk is nearly supersymmetric.

We also showed that our quintessence model can evade the severe bounds coming from fifth-force
experiments even if we are dealing with a scalar. The reason is the fact that the rigid four-cycle supporting the
Standard Model does not intersect the fibre divisor, and so the quintessence scalar
couples to ordinary matter only via its mixing with the volume mode which, in turn, induces its mixing
with the SM cycle. However due to the fact that the fibre modulus is a flat direction at leading order
which is lifted only at subleading order, the mixing between $\tau_f$ and $\vo$ is suppressed,
resulting in a weaker-than-Planckian coupling of the quintessence field to SM particles.

On the other hand, if dark matter is realised in terms of quasi-stable Kaluza-Klein states,
direct couplings between dark energy and dark matter are allowed,
leading to a scenario of coupled quintessence within string theory.
We studied in detail the dynamics of the quintessence scalar in our set-ups, investigating the
nature of fixed points and the late time evolution of dark energy,
showing that the main features of our scenario make it compatible with observations.

To conclude, our model explicitly shows that concrete string theory constructions are able
to lead to scalar theories of quintessence that, besides generating naturally flat scalar potentials,
also provide couplings to Standard Model matter that are weak enough to satisfy present constraints on fifth forces.

\acknowledgments
We would like to thank Cliff Burgess, Fernando Quevedo and Joseph Conlon for useful discussions.
GT is supported by an STFC Advanced Fellowship ST/H005498/1. FGP is supported by Funda\c{c}\~{a}o para a Ci\^{e}ncia e  Tecnologia (Portugal) through the grant SFRH/BD/35756/2007 and by the Theoretical Physics Department of the University of Oxford.

\begin{appendix}

\section{Canonical normalisation}
\label{AppendixA}

In this section we compute the canonical normalisation for the fields involved, using the well-known techniques developed in \cite{Conlon:2007gk,Conlon:2006tj,Reheating,LVScouplings}.
The Lagrangian for the quadratic field fluctuations around their minima, $\tau_i \,=\,\langle \tau_i\rangle +\delta \tau_i$, takes the form:
\be
{\cal L}\,=\,K_{i j} \partial_\mu \delta \tau_i  \partial^\mu \delta \tau_i -\langle V\rangle -\frac12  V_{i j} \delta \tau_i \delta \tau_j\,,
\ee
where $K_{ij}$ is the K\"ahler matrix, and $V_{ij}$ the matrix of second derivatives of the potential.
The leading terms in the K\"ahler matrix $K_{ij}$ look like:
\be
 K_{ij}\simeq\left(
\begin{array}{ccc}
 \frac{3}{8 \tau_f^2} & -\frac{1}{4 \vo \tau_f} & -\frac{3 \sqrt{\tau _s}}{8 \vo  \tau _f}  \\
 -\frac{1}{4 \vo  \tau _f} & \frac{1}{2 \vo ^2} & 0  \\
 -\frac{3 \sqrt{\tau _s}}{8 \vo  \tau _f} & 0 & \frac{3}{8 \vo  \sqrt{\tau _s}}
\end{array}
\right),
\label{eq:KahlerMetric1}
\ee
where we used $(\tau_f,\vo,\tau_s)$ as coordinates of the K\"ahler moduli space. We write the original moduli
in terms of the canonically normalised fields around their minima as:
\be\label{canfi}
\left(
\begin{array}{c}
 \delta \tau_f \\ \delta \vo \\ \delta \tau_s
 \end{array}
 \right)
\,=\,\mc{C}\left(
\begin{array}{c}
 \delta \phi \\ \delta \chi_{\vo} \\ \delta \chi_s
 \end{array}
 \right),
\ee
where $\mc{C}$ is the matrix that diagonalises both the kinetic and the mass terms. It can be written as:
\be
\mc{C}=\left(
\begin{array}{ccc}
\vec{v}_{(1)} & \vec{v}_{(2)}&\vec{v}_{(3)}
\end{array}
\right),
\ee
where $\vec{v}_{(i)}$ are the eigenvectors of the mass matrix $M_{ik}\,=\,\frac12 \,K^{-1}_{ij} V_{jk}$,
normalised such that $\mc{C}^T_{ik}\,  K_{km}\,\mc{C}_{mj}\,=\,\delta_{ij}$. The mass matrix reads explicitly:
\be
M_{ij}\,=\,\frac12 \left(
\begin{array}{ccc}
 -\frac{4 {\tau_f} ({f_8}+2 {f_6} {\tau_f}+2 {f_7} {\tau_s}) \vo}{\vo^{3+p}\left(3 {\tau_s}^{3/2}-2 \vo\right)} & -\frac{4 ({f_3}+2 {f_4}) {\tau_f} \sqrt{{\tau_s}}}{\left(3 {\tau_s}^{3/2}-2 \vo\right) \vo^3} & -\frac{4 ({f_4}+2 {f_5}) {\tau_f}}{3 {\tau_s}^2 \vo^2-2 \sqrt{{\tau_s}} \vo^3} \\
 -\frac{2 \vo \left(2 ({f_6} {\tau_f}+{f_7} {\tau_s}) \vo+3 {f_8} \left(-{\tau_s}^{3/2}+\vo\right)\right)}{\vo^{3+p}
 \left(3 {\tau_s}^{3/2}-2 \vo\right)} & \frac{6 {f_3} {\tau_s}^2-2 (3 {f_3}+2 {f_4}) \sqrt{{\tau_s}} \vo}{\left(3 {\tau_s}^{3/2}-2 \vo\right) \vo^3} & \frac{6 {f_4} \left({\tau_s}^{3/2}-\vo\right)-4 {f_5} \vo}{3 {\tau_s}^2 \vo^2-2 \sqrt{{\tau_s}} \vo^3} \\
 -\frac{4 \left(3 ({f_8}+2 {f_6} {\tau_f}) {\tau_s} \vo+4 {f_7} \sqrt{{\tau_s}} \vo^2\right)}{\vo^{3+p}\left(9 {\tau_s}^{3/2}-6 \vo\right)} & \frac{4 \left(3 {f_3} {\tau_s}^{3/2}+4 {f_4} \vo\right)}{3 \vo^3 \left(-3 {\tau_s}^{3/2}+2 \vo\right)} & -\frac{4 \left(3 {f_4} {\tau_s}^{3/2}+4 {f_5} \vo\right)}{3 \vo^2 \left(3 {\tau_s}^{5/2}-2 {\tau_s} \vo\right)}
\end{array}
\right).
\label{massma}
\ee
The quantities $f_i$ are independent from the volume at leading order in an inverse volume expansion. Their expressions are quite long, and we relegate them in appendix \ref{app-B}. In writing the previous mass matrix, we added to the leading part of the potential, eq. (\ref{eq:Vleading}),
the subleading contribution due to polyinstanton corrections, eq. (\ref{eq:Vsubleading0}).

Notice that, as expected, the first column of the mass matrix is suppressed by a factor $1/\vo^{3+p}$, since it is  due to the subleading
poly-instanton potential. In other words, the first column of the matrix (\ref{massma}) would be zero in the absence of the subleading poly-instanton contribution.

It is straightforward, but long, to extract the eigenvalues and the eigenvectors of the mass matrix of eq. (\ref{massma}),
working at leading order in an expansion in inverse powers of the volume.
The expressions of the eigenvectors turn out to be particularly  complicated. It is convenient to present them
at leading order in an expansion in the quantity $\tau_s$, that we can assume to be small with respect to the other quantities. We find:
\be\label{eigval}
\mc{C}\,=\,
\left(
\begin{array}{ccc}
 \frac{2{\tau_f}}{\sqrt{3}} & \sqrt{\frac{2}{3}}{\tau_f} & 0 \\
 \frac{c_0}{\vo^{p-1}} & \sqrt{\frac{3}{2}} \vo & - {\tau_s}^{3/4} \sqrt{3 \vo} \\
 \frac{c_1}{\vo^p} & \sqrt{6}{\tau_s} & \frac{2{\tau_s}^{1/4} \sqrt{\vo}}{\sqrt{3}}
\end{array}
\right).
\ee
The two last entries in the first column  are suppressed by powers of the volume, and are due to the effect of the poly-instanton corrections.
The quantities $c_0$ and $c_1$ are independent from the volume (at leading order in a inverse volume expansion) and read:
\bea
c_0&=&-\frac{\tau_s^{\frac{p}{2}-1}}{2 e \pi  {W_0}}3^{-\frac{3}{2}+p} (-2 a A {C_1} \pi +b B (c (-{C_1}+{C_2})+2 {C_1} \pi ))\left(\frac{{W_0}}{
b B-a A}\right)^{1+p},
\\
c_1&=&\frac{4\,\tau_s^{\frac{p}{2}}\, 3^{-\frac{3}{2}+p} (2 a A {C_1} \pi +b B (c ({C_1}-{C_2})-2 {C_1} \pi ))  \left(\frac{{W_0}}{-a A+b B}\right)^p}{(a A-b B) e \pi }.
\eea
The lightest eigenvalue of the mass matrix scales as $m_{\phi}^2\,\propto\,1/\vo^{3+p}$, and provides the effective mass for the quintessence field in the poly-instanton dominated region of the quintessence potential (region II of figure \ref{fig:Potential}).

In this region, the volume and $\tau_s$ are stabilised at their minima since their mass is very large. From eqs (\ref{canfi}) and (\ref{eigval}), one obtains that  the displacement of  the light field $\tau_f$ from its minimum can be expressed in terms of its canonical counterpart as:
\be\label{cant1}
\delta \tau_f\,\simeq\,\frac{2}{\sqrt{3}}\,\tau_f\,\delta \phi \hskip1cm\Rightarrow\hskip1cm \tau_f\,\simeq\, e^{\frac{2}{\sqrt{3}} \,\phi},
\ee
justifying the canonical normalisation we used in eq. (\ref{eq:VCanNorm}).  The diagonalising matrix (\ref{eigval}), when
compared with  (\ref{canfi}),  also tells us that, within the approximations we are using,
 we can write the modulus $\tau_s$ in terms of canonically normalised fields as:
\be\label{expt3}
\frac{\delta \tau_s}{\langle\tau_s\rangle}\,\simeq\, \frac{c_1}{\vo^p}\,\delta \phi+
\sqrt{6}\,\delta \chi_{\vo}+\frac{2 \sqrt{\vo}}{\sqrt{3}\langle\tau_s\rangle^{3/4}}\, \delta \chi_s\,,
\ee
reproducing the scalings of eq. (\ref{eq:canNorm3}).

\section{Explicit form of the coefficients of the mass matrix}
\label{app-B}

Here we collect the complete expressions for the quantities $f_i$ appearing in the mass matrix $M_{ij}$, eq. (\ref{massma})
(we use the same conventions introduced in our section \ref{QuintPot}):
\bea
f_1&=&\frac{((a - b) b B \tau_s + Z - a \tau_s Z)}{(4 (a - b) b B \tau_s + Z - 4 a \tau_s Z)} \nn \\
f_2&=&\frac{3 \sqrt{{\tau_s}} {W_0} \left(b^2 B {\tau_3}-Z+a {\tau_s} (-b B+Z)\right)}{Z \left(4 b^2 B {\tau_s}-Z+4 a {\tau_s} (-b B+Z)\right)}
 \nn \\
f_3&=&\frac{8}{3} \left(-36 (-1+{f_1}) {f_1} {\tau_s} {W_0}^2-9 {f_2} \sqrt{{\tau_s}} {W_0} Z+2 {f_2}^2 Z^2\right) \nn \\
f_4&=&\frac{4}{3} {f_2} \left(-{f_2} Z (4 (a-b) b B {\tau_s}+Z-4 a {\tau_s} Z)+6 \sqrt{{\tau_s}} {W_0} ((a-b) b B {\tau_s}+Z-a {\tau_s} Z)\right) \nn \\
f_5&=&\frac{2}{3} {f_2} \Big[-6 {\tau_s}^{3/2} {W_0} (b (-a+b) B (-2+(a+b) {\tau_s})+a (-2+a {\tau_s}) Z)+{f_2} \Big(8 (a-b)^2 b^2 B^2 {\tau_s}^2 \nonumber\\
&& +8 b (-a+b) B {\tau_s} (-1+3 a {\tau_s}+b {\tau_s}) Z+(-1+8 a {\tau_s} (-1+2 a {\tau_s})) Z^2\Big)\Big] \nn
\\
f_6&=&
-\frac{16 \pi{f_2}^{1+p}}{3 e} \Big[ 3 b B (-c ({C_1}-{C_2}) (-1+\pi  {\tau_f})+(-a {C_1}+b {C_2}) \pi  {\tau_s}) {W_0} \nonumber\\
&& +\pi  \left(4 B (a {C_1}+b ((-1+b) {C_1}-b {C_2})) {f_2} \sqrt{{\tau_s}}+3 {C_1} (-2+2 \pi  {\tau_f}+a {\tau_s}) {W_0}\right) Z-4 a {C_1} {f_2} \pi  \sqrt{{\tau_s}} Z^2\Big] \nn
\\
f_7&=&\frac{4 {f_2}^{1+p}}{3 e \sqrt{{\tau_s}}} \Big[
3 b B \sqrt{{\tau_s}} \left(-2 a {C_1} \pi -2 b^2 {C_2} \pi  {\tau_s}+b (c ({C_1}-{C_2}) (-1+2 \pi  {\tau_f})+2 \pi  ({C_2}+a {C_1} {\tau_s}))\right) {W_0}\nonumber\\
&& -2 \pi  \left(2 B (a {C_1}+b ((-1+b) {C_1}-b {C_2})) {f_2} (-1+4 b {\tau_s})+3 {C_1} \sqrt{{\tau_s}} (b (-1+2 \pi  {\tau_f})+a (-1+b {\tau_s})) {W_0}\right) Z\nonumber\\
&&+4 a {C_1} {f_2} \pi  (-1+4 b {\tau_s}) Z^2\Big] \nn \\
f_8&=&\frac{8 {f_2}^{1+p}}{3 e} \Big[
3 b B (c ({C_1}-{C_2}) (-1+2 \pi  {\tau_f})+2 (a {C_1}-b {C_2}) \pi  {\tau_s}) {W_0} \nn \\
&&
-2 \pi  \left(2 B (a {C_1}+b ((-1+b) {C_1}-b {C_2})) {f_2} \sqrt{{\tau_s}}+3 {C_1} (-1+2 \pi  {\tau_f}+a {\tau_s}) {W_0}\right) Z+4 a {C_1} {f_2} \pi  \sqrt{{\tau_s}} Z^2\Big]\nonumber
\eea

\end{appendix}


\begin{thebibliography}{10}


\bibitem{QuintReview}
  E.~J.~Copeland, M.~Sami and S.~Tsujikawa,
  Int.\ J.\ Mod.\ Phys.\ D {\bf 15} (2006) 1753
  [hep-th/0603057].

\bibitem{Kolda:1998wq}
  C.~F.~Kolda and D.~H.~Lyth,
  Phys.\ Lett.\ B {\bf 458} (1999) 197
  [hep-ph/9811375].

\bibitem{Adelberger:2003zx}
  E.~G.~Adelberger, B.~R.~Heckel and A.~E.~Nelson,
  Ann.\ Rev.\ Nucl.\ Part.\ Sci.\  {\bf 53} (2003) 77
  [arXiv:hep-ph/0307284].

\bibitem{Carroll:1998zi}
  S.~M.~Carroll,
  Phys.\ Rev.\ Lett.\  {\bf 81} (1998) 3067
  [astro-ph/9806099].


\bibitem{Choi:1999xn}
  K.~Choi,
Phys.\ Rev.\ D {\bf 62} (2000) 043509  [hep-ph/9902292].  

\bibitem{Kaloper:2008qs}
  N.~Kaloper and L.~Sorbo,
  Phys.\ Rev.\ D {\bf 79} (2009) 043528
  [arXiv:0810.5346 [hep-th]].

\bibitem{Panda:2010uq}
  S.~Panda, Y.~Sumitomo and S.~P.~Trivedi,
  Phys.\ Rev.\ D {\bf 83} (2011) 083506  [arXiv:1011.5877 [hep-th]].  

\bibitem{Gupta:2011yj}
  G.~Gupta, S.~Panda and A.~A.~Sen,
  Phys.\ Rev.\ D {\bf 85} (2012) 023501  [arXiv:1108.1322 [astro-ph.CO]].  


\bibitem{Cicoli:2011ct}
  M.~Cicoli, F.~G.~Pedro and G.~Tasinato,
  JCAP {\bf 1112} (2011) 022  [arXiv:1110.6182 [hep-th]].



\bibitem{ADDstrings}
  M.~Cicoli, C.~P.~Burgess and F.~Quevedo,
  JHEP {\bf 1110} (2011) 119  [arXiv:1105.2107 [hep-th]].  

\bibitem{SLED}
  Y.~Aghababaie, C.~P.~Burgess, S.~L.~Parameswaran and F.~Quevedo,
  Nucl.\ Phys.\  B {\bf 680} (2004) 389
  [arXiv:hep-th/0304256];

  Y.~Aghababaie {\it et al.},
  JHEP {\bf 0309} (2003) 037
  [arXiv:hep-th/0308064];

  J.~Vinet, J.~M.~Cline,
  Phys.\ Rev.\  {\bf D71 } (2005)  064011
  [hep-th/0501098].

\bibitem{Albrecht:2001xt}
  A.~Albrecht, C.~P.~Burgess, F.~Ravndal and C.~Skordis,
  Phys.\ Rev.\ D {\bf 65} (2002) 123507
  [astro-ph/0107573].

\bibitem{Burgess:2004ib}
  C.~P.~Burgess,
  AIP Conf.\ Proc.\  {\bf 743} (2005) 417
  [hep-th/0411140].

\bibitem{Nilles:2003km}
  H.~-P.~Nilles, A.~Papazoglou and G.~Tasinato,
  Nucl.\ Phys.\ B {\bf 677} (2004) 405
  [hep-th/0309042].

\bibitem{Burgess:2011mt}
  C.~P.~Burgess and L.~van Nierop,
  JHEP {\bf 1104} (2011) 078
  [arXiv:1101.0152 [hep-th]].

\bibitem{FibredCY}
  M.~Cicoli, M.~Kreuzer and C.~Mayrhofer,
  JHEP {\bf 1202} (2012) 002  [arXiv:1107.0383 [hep-th]].  

\bibitem{LVS}
  V.~Balasubramanian, P.~Berglund, J.~P.~Conlon and F.~Quevedo,
  JHEP {\bf 0503} (2005) 007
  [arXiv:hep-th/0502058].

\bibitem{Blumenhagen:2007sm}
  R.~Blumenhagen, S.~Moster and E.~Plauschinn,
  JHEP {\bf 0801} (2008) 058
  [arXiv:0711.3389 [hep-th]].

\bibitem{Cicoli:2011qg}
  M.~Cicoli, C.~Mayrhofer and R.~Valandro,
  JHEP {\bf 1202} (2012) 062  [arXiv:1110.3333 [hep-th]].


\bibitem{Kachru:2003aw}
  S.~Kachru, R.~Kallosh, A.~D.~Linde and S.~P.~Trivedi,
  Phys.\ Rev.\ D {\bf 68} (2003) 046005
  [hep-th/0301240].

\bibitem{bkq}
  C.~P.~Burgess, R.~Kallosh and F.~Quevedo,
  JHEP {\bf 0310} (2003) 056
  [hep-th/0309187].

\bibitem{Cicoli:2012fh}
  M.~Cicoli, A.~Maharana, F.~Quevedo and C.~P.~Burgess,
  arXiv:1203.1750 [hep-th].

\bibitem{StringLoops}
  M.~Cicoli, J.~P.~Conlon and F.~Quevedo,
  JHEP {\bf 0801} (2008) 052
  [arXiv:0708.1873 [hep-th]].
  M.~Cicoli, J.~P.~Conlon and F.~Quevedo,
  JHEP {\bf 0810} (2008) 105
  [arXiv:0805.1029 [hep-th]].

\bibitem{Blumenhagen:2008ji}
  R.~Blumenhagen and M.~Schmidt-Sommerfeld,
  JHEP {\bf 0807} (2008) 027
  [arXiv:0803.1562 [hep-th]].

\bibitem{Conlon:2008cj}
  J.~P.~Conlon, R.~Kallosh, A.~D.~Linde and F.~Quevedo,
  JCAP {\bf 0809} (2008) 011
  [arXiv:0806.0809 [hep-th]].

\bibitem{Cicoli:2012cy}
  M.~Cicoli, G.~Tasinato, I.~Zavala, C.~P.~Burgess and F.~Quevedo,
  arXiv:1202.4580 [hep-th].

\bibitem{Conlon:2006tj}
  J.~P.~Conlon, D.~Cremades and F.~Quevedo,
  JHEP {\bf 0701} (2007) 022
  [arXiv:hep-th/0609180].

\bibitem{Copeland:1997et}
  E.~J.~Copeland, A.~R.~Liddle and D.~Wands,
  Phys.\ Rev.\  D {\bf 57} (1998) 4686
  [arXiv:gr-qc/9711068].

\bibitem{Hellerman}
  S.~Hellerman, N.~Kaloper and L.~Susskind,
JHEP {\bf 0106} (2001) 003  [hep-th/0104180].  

\bibitem{Fischler}
  W.~Fischler, A.~Kashani-Poor, R.~McNees and S.~Paban,
  JHEP {\bf 0107} (2001) 003
  [arXiv:hep-th/0104181].

\bibitem{Amendola}
  L.~Amendola,
  Phys.\ Rev.\ D {\bf 62} (2000) 043511  [astro-ph/9908023].  

\bibitem{CoupledDEDM}
  C.~G.~Boehmer, G.~Caldera-Cabral, R.~Lazkoz and R.~Maartens,
  Phys.\ Rev.\ D {\bf 78} (2008) 023505
  [arXiv:0801.1565 [gr-qc]].

\bibitem{Conlon:2007gk}
  J.~P.~Conlon and F.~Quevedo,
  JCAP {\bf 0708} (2007) 019
  [arXiv:0705.3460 [hep-ph]].

\bibitem{Reheating}
  M.~Cicoli and A.~Mazumdar,
  JCAP {\bf 1009} (2010) 025  [arXiv:1005.5076 [hep-th]].  

\bibitem{LVScouplings}
  L.~Anguelova, V.~Calo and M.~Cicoli,
  JCAP {\bf 0910} (2009) 025  [arXiv:0904.0051 [hep-th]].  


\end{thebibliography}
\end{document}